\documentclass[12pt,draftclsnofoot,letterpaper,oneside,onecolumn]{IEEEtran}
\def\BibTeX{{\rm B\kern-.05em{\sc i\kern-.025em b}\kern-.08em
    T\kern-.1667em\lower.7ex\hbox{E}\kern-.125emX}}
\usepackage{amsmath}
\usepackage{amsthm}
\usepackage{amsfonts}
\usepackage{amssymb}
\usepackage{graphicx}
\usepackage{cite}
\usepackage{subfigure}
\usepackage{epstopdf}
\usepackage{balance}
\usepackage{algorithm}
\usepackage{algorithmic}
\usepackage{enumerate}
\usepackage{color}
\usepackage{array}
\usepackage{indentfirst}
\usepackage{multirow}
\usepackage{booktabs}
\usepackage{color}
\usepackage{slashbox}
\usepackage{diagbox}
\usepackage{threeparttable}
\usepackage[svgnames,table]{xcolor}
\usepackage{bm}

\makeatletter

\newcommand{\Rmnum}[1]{\expandafter\@slowromancap\romannumeral #1@}
\makeatother

\begin{document}
\title{ Deep Transfer Learning for Signal Detection in Ambient Backscatter Communications }

\author{\IEEEauthorblockN{Chang Liu, \emph{Member, IEEE}, Zhiqiang Wei, \emph{Member, IEEE}, \\ Derrick Wing Kwan Ng, \emph{Senior Member, IEEE}, \\ Jinhong Yuan, \emph{Fellow, IEEE}, and Ying-Chang Liang, \emph{Fellow, IEEE}} 

\thanks{C. Liu, Z. Wei, D. W. K. Ng, and J. Yuan are with the School of Electrical Engineering and Telecommunications, the University of New South Wales, Australia (email: chang.liu19@unsw.edu.au; zhiqiang.wei@unsw.edu.au; w.k.ng@unsw.edu.au; j.yuan@unsw.edu.au).
}

\thanks{Y.-C. Liang is with the Center for Intelligent Networking and Communications (CINC), University of Electronic Science and Technology of China (UESTC), Chengdu, China (e-mail: liangyc@ieee.org).}

\thanks{This work has been accepted in part to present at the IEEE Global Communication Conference (Globecom), 2020 \cite{Liu2020con}.}

}

%
%
%

\maketitle

\vspace{-1.2cm}

\begin{abstract}
Tag signal detection is one of the key tasks in ambient backscatter communication (AmBC) systems. However, obtaining perfect channel state information (CSI) is challenging and costly, which makes AmBC systems suffer from a high bit error rate (BER). To eliminate the requirement of channel estimation and to improve the system performance, in this paper, we adopt a deep transfer learning (DTL) approach to implicitly extract the features of channel and directly recover tag symbols. To this end, we develop a DTL detection framework which consists of offline learning, transfer learning, and online detection. Specifically, a DTL-based likelihood ratio test (DTL-LRT) is derived based on the minimum error probability (MEP) criterion. As a realization of the developed framework, we then apply convolutional neural networks (CNN) to intelligently explore the features of the sample covariance matrix, which facilitates the design of a CNN-based algorithm for tag signal detection. Exploiting the powerful capability of CNN in extracting features of data in the matrix formation, the proposed method is able to further improve the system performance. In addition, an asymptotic explicit expression is also derived to characterize the properties of the proposed CNN-based method when the number of samples is sufficiently large. Finally, extensive simulation results demonstrate that the BER performance of the proposed method is comparable to that of the optimal detection method with perfect CSI.
\end{abstract}

\vspace{-0.1cm}

\begin{IEEEkeywords}
\vspace{-0.1cm}
Ambient backscatter communications, signal detection, deep transfer learning, Internet-of-Things.
\end{IEEEkeywords}

\newpage
\section{Introduction\label{sect: intr}}
The Internet-of-Things (IoT), which enables the connections of massive devices through the internet, is one of the key technologies of the fifth-generation (5G) wireless communications \cite{mei2019survey, wong2017key}. With the rapid development of wireless sensor technologies, a huge number of devices are connected via the IoT. However, the power consumption of the associated wireless devices becomes a critical problem \cite{palattella2016internet, chen2020massive, zhang2020prospective}. For example, most of IoT devices are expected to be powered by batteries with limited energy storage capacity and lifetime. As a result, various advanced technologies have been proposed in the literature to tackle the issues \cite{wu2017overview}.
One of the promising solutions is ambient backscatter communication (AmBC), a low-cost and energy-efficient communication scheme which enables passive backscatter devices (e.g., tags, sensors) to transmit their information bits to readers over ambient radio-frequency (RF) signals (e.g., Wi-Fi signals, cellular base station signals, and TV tower radio signals) \cite{van2018ambient, yang2017modulation}. Since ambient RF signals always exist in modern cities, the communication between passive devices is possible without requiring additional power for generating a carrier wave \cite{xie2014managing}. In particular, an AmBC tag of an AmBC system could transmit its binary tag symbols by choosing whether to backscatter the ambient RF signals or not. Thus, the main task of an AmBC system is to perform tag signal detection, i.e., recovering the tag signal at the reader, which has attracted tremendous attention from both academia and industry \cite{van2018ambient}, respectively.
Generally, there are two main challenges for tag detection: (1) since both the direct link signal from the RF source and the backscatter link signal from the tag could be received by the reader simultaneously, the received direct link signal generally causes severe interference to the received backscatter link signal; (2) in contrast to the traditional wireless communication systems, estimating the channel state information (CSI) in AmBC systems is challenging due to the lack of pilot signals sent from the ambient RF source.

Recently, various effective algorithms and schemes have been proposed for tag signal detection in AmBC systems. For example, Liu \emph{et al.} \cite{liu2013ambient} implemented practical ambient backscatter devices and proposed an energy detector adopting a differential coding scheme to decode the tag signals, which paves the way for realizing AmBC systems. Lu \emph{et al.} \cite{lu2015signal} developed an improved energy detection method which however requires the knowledge of perfect CSI. To overcome this problem, Qian \emph{et al.} \cite{qian2017semi} designed a semi-coherent detection method requiring only a few pilots and unknown data symbols, which substantially reduces the signaling overhead and decoding computational complexity.
Furthermore, in order to eliminate the process of channel estimation, Wang \emph{et al.} \cite{wang2016ambient} adopted a differential encoding scheme on tag signals and proposed a minimum BER detector with the corresponding optimal detection threshold. On top of \cite{wang2016ambient}, Qian \emph{et al.} in \cite{qian2016noncoherent} provided fundamental studies of the BER performance for non-coherent detectors.
Nevertheless, the above methods either require channel estimation explicitly or lead to an unsatisfactory detection performance. As a remedy, machine learning (ML)-based methods have been proposed recently which aim to directly recover the tag signals with only a few (training) pilots without the need of estimating relevant channel parameters explicitly. For example, Zhang \emph{et al.} \cite{zhang2018constellation} proposed a clustering method to extract the features of constellation symbols and designed two constellation learning-based signal detection methods, which achieve satisfying BER performance. However, these proposed methods were designed for systems when the ambient RF source signal is a constellation modulated signal.
Furthermore, Hu \emph{et al.} \cite{hu2019machine} transformed the task of tag signal detection into a classification task and designed the support vector machine (SVM)-based energy detection method to improve the BER performance.
However, the proposed ML-based method requires a large number of training pilots and there exists a large gap between the proposed method and the optimal method.
In contrast, different from traditional ML methods, deep learning technology, which adopts a neural network to intelligently explore features in a data-driven manner, has been shown to be able to achieve outstanding performance in many areas, such as natural language processing \cite{young2018recent}, computer vision \cite{goodfellow2016deep}, and also wireless communications \cite{liu2020deepresidualL, liu2020deep, liu2020location, yuan2020learning, liu2020deepresidual}. Nevertheless, practical wireless channels change over time with a huge dynamic range. Besides, transmission in wireless systems can be bursty and the transmission pattern is time varying.
In this case, a well-trained deep neural network (DNN) can hardly perform well in the detection tasks under different channel coefficients.
To overcome these limitations, a more practical approach which can dynamically adjust the weights in DNNs adapting the changing of channel environment is expected.

In contrast to the traditional ML methods, deep transfer learning (DTL) is proposed to adopt a DNN to extract the time-varying features with a few online training data by transferring knowledge from a source domain to a target domain \cite{pan2009survey, weiss2016survey, tan2018survey}.
{However, there are two main challenges when applying DTL to signal detection in AmBC systems:
(1) since the received signal power from the backscatter link is much weaker than that from the direct link, it leads to a small difference in the two hypotheses (i.e., the tag chooses to backscatter the ambient RF source signals or not), which brings difficulties to DTL in completing the classification task \cite{goodfellow2016deep};
(2) different from the general classification tasks in the field of computer vision in which an error rate of around $10\%$ is acceptable \cite{goodfellow2016deep}, the classification task in AmBC systems, i.e., the tag signal detection task, has a much lower bit error rate (BER) requirement, which is generally lower than $10^{-2}$ \cite{3GPPTS2018Quality, van2018ambient}.}
Motivated by this, in this paper, we try to overcome these challenges and propose a framework based on the DTL approach for tag signal detection, which adopts a DNN to transfer the knowledge learned from one tag detection task under the offline channel coefficients to another different but related tag detection task in real-time. The proposed framework can adaptively fine-tune the detector for different channel environments to further improve the detection performance.
To our best knowledge, this is the first study applying DTL to tag signal detection for AmBC systems.
The main contributions of this paper are as follows:
\begin{enumerate}[(1)]
\item We introduce a universal DTL-based tag signal detection framework which consists of offline learning, transfer learning, and online detection. Different from conventional methods which require explicit channel estimation, our proposed scheme adopts a DNN to extract the features of channel and directly recover the tag symbols, providing a more practical approach for tag signal detection.
    Specifically, according to the minimum error probability (MEP) criterion, a DTL-based likelihood ratio test (LRT) is derived for tag signal detection, which enables the design of an effective detector.
\item Under the developed framework, we creatively adopt a convolutional neural network (CNN) to explore the features of the sample covariance matrix. Inspired by the novel covariance matrix aware-CNN (CM-CNN) structure \cite{liu2019deep}, \cite{xie2019activity}, we develop a DTL-oriented covariance matrix aware neural network (CMNet) for tag signal detection and propose the related tag signal detection algorithm. Exploiting the powerful capability of CNN in exploring features of data in a matrix form, the proposed method could extract more discriminative features to further improve the BER performance.
\item Although it is intractable to directly analyze the performance of a CNN, we formulate the proposed CMNet as a non-linear function and derive an asymptotic explicit expression to characterize its properties when the number of samples is sufficiently large.
\item Extensive experiments have been conducted using both the modulated and complex Gaussian ambient sources. The results show that the proposed algorithm can achieve an outstanding  BER performance which is close to that of the optimal LRT method with perfect CSI.
\end{enumerate}

The remainder of this paper is organized as follows. Section \Rmnum{2} formulates the AmBC system model. In Section \Rmnum{3}, we develop a DTL-based detection framework for the AmBC system. As a realization of the developed framework, Section \Rmnum{4} proposes a novel CMNet-based tag signal detection algorithm and provides the corresponding theoretical analysis. Extensive simulation results are presented in Section \Rmnum{5}, and Section \Rmnum{6} finally concludes the work of this paper.

The notations used in this paper are shown in the following. The superscripts $T$ and $H$ indicate the transpose and conjugate transpose, respectively. Term ${\mathcal{CN}}( \bm{\mu},\mathbf{\Sigma} )$ represents a circularly symmetric complex Gaussian (CSCG) distribution with a mean vector $\bm{\mu}$ and a covariance matrix $\mathbf{\Sigma}$. Term ${\bf{I}}_M$ is used to denote the $M$-by-$M$ identity matrix and ${\mathbf{0}}$ is used to denote the zero vector. $(\cdot)^{-1}$ indicates the matrix inverse operation. $\max(a,b)$ is the maximum value of $a$ and $b$. Subscripts $S$ and $T$ represent the source domain and the target domain, respectively. $|\cdot|$ denotes the cardinality of a set. $\mathrm{Re}(\cdot)_{i,j}$ and $\mathrm{Im}(\cdot)_{i,j}$ represent the $i$-th row and the $j$-th column elements of the real part and imaginary part, respectively. $\det(\cdot)$ is the determinant operator. $P(\cdot)$ and $E(\cdot)$ represent the probability of an event and the statistical expectation, respectively. $\|\cdot\|^2$ denotes the norm of an input vector. $\exp(\cdot)$ represents the exponential function.

\begin{figure}[t]
  \centering
  \includegraphics[width=0.5\linewidth]{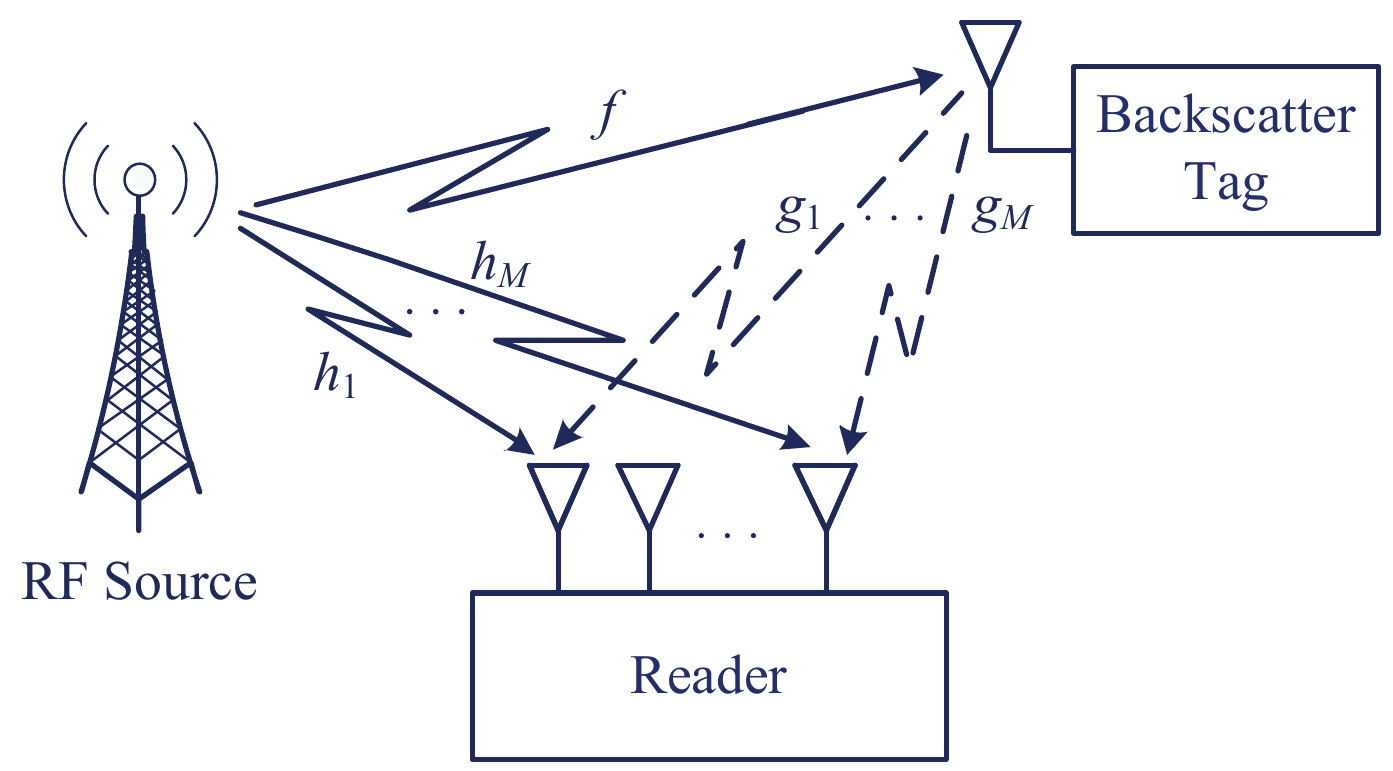}\vspace{-0.6cm}
  \caption{ An illustration of the considered ambient backscatter communication system. }\vspace{-1cm}
\end{figure}

\section{System Model}
In this paper, we consider a general AmBC system, which consists of an ambient RF source, a tag, and a reader, as depicted in Fig. 1. Both the RF source and the passive tag are equipped with a single antenna, while the reader is equipped with an $M$-element antenna array for signal detection. Due to the broadcasting nature of the RF source, the transmitted RF signal is received by both the reader and the tag simultaneously. Although the tag is a passive device, it can transmit its binary tag symbols by choosing whether to reflect the ambient RF signals to the reader. In this case, the reader can then decode the tag symbols through sensing the changes from the received signals.
The frame structure of the signal model at the reader is shown in Fig. 2, which consists of $P$ pilot symbols{{\footnotemark}}\footnotetext{{Note that the pilots are not used for acquiring the channel information explicitly, while they are exploited to build the online training dataset for the learning algorithm to capture the features of the real-time channel to facilitate the tag signal detection. By doing this, the proposed framework not only eliminates the requirement of explicit channel estimation, but also further improves the system performance.}} and $T-P$ $(T>P)$ data symbols in one frame. For the pilots, the tag symbols are known by the reader, and the remaining tag symbols are used for data transmission.
{The tag frame structure considered in this paper is designed for slow fading channel models, i.e., the channel remains unchange during each frame, which is commonly adopted in the literature for AmBC and RFID applications, e.g., \cite{van2018ambient, finkenzeller2010rfid}. For the case of fast fading channel, since the pilot-based training data and the test data come from different channel environments, the learned features by the neural network from the training data do not match with the test data and thus they cannot be used for the tag signal detection. As a result, a new tag frame structure is desired which will be left for future work.} In addition, the tag transmits its bits at a rate $N$ times lower than the sampling rate of the RF source signal. Thus, the source-to-tag ratio (STR) which is defined as the number of RF source symbols in one tag symbol period is $N$.
In other words, each tag symbol remains the same within the $N$ RF source symbol periods.

\begin{figure}[t]
  \centering
  \includegraphics[width=0.6\linewidth]{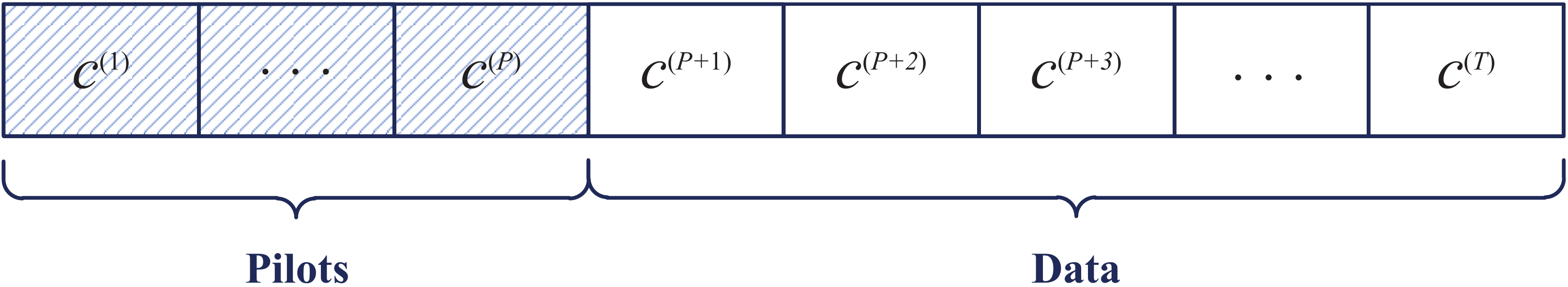}\vspace{-0.6cm}
  \caption{ The tag frame structure of the considered AmBC system. The shaded parts are the pilot signals labelled and the remaining parts are the test data.}\vspace{-1cm}
\end{figure}

Denote by $c^{(t)}\in \mathcal{C} = \{0,1\}$ the $t$-th tag symbol with the binary on-off keying modulation, i.e., $c^{(t)} = 0$ refers to that the tag does not reflect the RF source signal; otherwise, the tag chooses to reflect the RF source signal. Correspondingly, we use $s_n^{(t)}$ to represent the $n$-th RF source signal sample within the tag symbol $c^{(t)}$.
Let $ \mathbf{x}_n^{(t)}={{[x_{n,1}^{(t)},x_{n,2}^{(t)},\cdots ,x_{n,M}^{(t)}]}^{T}}, n \in \{ 0,1,\cdots,N-1\}$, represent the $n$-th observation vector within the $t$-th, $t\in \{ 1,\cdots,T \}$, tag symbol period, where $x_{n,m}^{(t)}$, $m \in \{1,2, \cdots, M\}$, denotes the $n$-th discrete-time sample observed at the $m$-th antenna element.
Therefore, the received signal at the reader can be expressed as \cite{guo2018exploiting, yang2015multi}
\begin{equation}\label{sr}
\mathbf{x}_n^{(t)} = \mathbf{h}s_n^{(t)} + \alpha f\mathbf{g}s_n^{(t)}c^{(t)} + \mathbf{u}_n^{(t)}, \forall n,t.
\end{equation}
Here, $\mathbf{h}=[h_1, h_2, \cdots, h_M ]^T$, where $h_m \in \mathbb{C}$ is the channel coefficient from the RF source to the $m$-th antenna at the reader. Similarly, $\mathbf{g}=[g_1, g_2, \cdots, g_M]^T$, where $g_m \in \mathbb{C}$ is the channel coefficient from the tag to the $m$-th antenna at the reader. Variables $f$, $\alpha\in \mathbb{C}$ are the channel coefficient from the RF source to tag and the reflection coefficient of the tag, respectively. In addition, $\mathbf{u}_n^{(t)}\in \mathbb{C}^{M\times1}$ is assumed to be an independent and identically distribution (i.i.d.) CSCG random noise vector with $\mathbf{u}_n^{(t)}\sim \mathcal{CN}( \mathbf{0},\sigma _u^2{{\mathbf{I}}_M} )$, where $\sigma_u^2$ represents the noise variance at each antenna of the reader. Considering a tag symbol is time invariant within $N$ RF source signal sampling periods, let
\begin{equation}\label{observation matrix} \vspace{-0.05cm}
\mathbf{X}^{(t)}=[\mathbf{x}_1^{(t)},\mathbf{x}_2^{(t)},\cdots,\mathbf{x}_N^{(t)}], \forall t, \vspace{-0.05cm}
\end{equation}
denote a sampling matrix collecting the $N$ observations of the $t$-th tag symbol at the reader. The task of the tag signal detection is to recover $c^{(t)}$ based on $\mathbf{X}^{(t)}$.
Thus, the tag signal detection can be further formulated as a binary hypothesis testing problem:
\begin{equation}\label{sensing model}
\begin{split}
 & {{H}_{1}}:\mathbf{x}_n^{(t)} = \mathbf{w}s_n^{(t)} + \mathbf{u}_n^{(t)}, \\
 & {{H}_{0}}:\mathbf{x}_n^{(t)} = \mathbf{h}s_n^{(t)} + \mathbf{u}_n^{(t)}, \\
\end{split}
\end{equation}
where $\mathbf{w} = \mathbf{h}+\alpha f\mathbf{g}$, and $H_1$ and $H_0$ denote the hypotheses that $c^{(t)}=1$ and $c^{(t)}=0$, respectively.

Note that the ambient source signal in practice may arise from either an unknown (or indeterminate) ambient RF source or a known and  discrete constellation ambient RF source \cite{qian2017semi}. Therefore, we consider two different ambient sources:
\begin{enumerate}[(a)]
\item \emph{Complex Gaussian Ambient Source}: $s_n^{(t)}$ is assumed to be a CSCG random variable{\footnotemark}\footnotetext{In the absence of any priori knowledge of $s_n^{(t)}$, a general approach is to model $s_n^{(t)}$ by a Gaussian random variable \cite{liu2019maximum, qian2017semi}.} with $s_n^{(t)}\sim \mathcal{C}\mathcal{N}(0,\sigma_s^2)$;
\item \emph{Modulated Ambient Source}: $s_n^{(t)}$ is assumed to be a Q-ary modulated symbol with power $\sigma_s^2$ drawn from a constellation set $\mathcal{S}=\{{S}_1,{S}_2, \cdots, {S}_Q\}$, with an equal probability.
\end{enumerate}
For the sake of presentation, we first define the received signal-to-noise ratio (SNR) of the direct link as
\begin{equation}\label{SNR}
\mathrm{SNR} = \frac{E(||\mathbf{h}s_n^{(t)}||^2)}{E(||\mathbf{u}_n^{(t)}||^2)}.
\end{equation}
Besides, the relative coefficient between the direct signal path and the backscattered signal path is defined as a ratio of their average channel gains which is given by
\begin{equation}\label{relative_SNR}
\zeta = \frac{E(||\alpha f \mathbf{g}||^2)}{E(||\mathbf{h}||^2)}.
\end{equation}

Then, in the following, we will introduce the optimal likelihood ratio test for above two cases and the corresponding results serve as a benchmark to the considered AmBC system.

\emph{1) \underline{Optimal LRT with Complex Gaussian Ambient Source:}}

When $s_n^{(t)}$ is a complex Gaussian signal as defined in (a), we have
\begin{equation}\label{random-x(n)}
\mathbf{x}_n^{(t)}\sim
\bigg\{ \begin{matrix}
   \mathcal{C}\mathcal{N}(\mathbf{0},\mathbf{\Sigma}_1),\;{{H}_{1}} \\
   \mathcal{C}\mathcal{N}(\mathbf{0},\mathbf{\Sigma}_0),\;{{H}_{0}}   \\
\end{matrix},
\end{equation}
where $\mathbf{\Sigma}_1 = \sigma_s^2\mathbf{w}\mathbf{w}^{H}+\sigma _{u}^{2}{{\mathbf{I}}_{M}}$ and $\mathbf{\Sigma}_0 = \sigma_s^2\mathbf{h}\mathbf{h}^{H}+\sigma _{u}^{2}{{\mathbf{I}}_{M}}$.
According to \cite{kay1998fundamentals}, if perfect CSI, e.g., $\mathbf{w}$ and $\mathbf{h}$, are known at the reader, we can then derive the logarithmic form of likelihood ratio test (LRT) under the complex Gaussian distributed ambient source:
\begin{equation}\label{L_R}
L_{\mathrm{CG}}({\mathbf{X}}^{(t)}) = \sum\limits_{n = 0}^{N - 1} \ln \left({\frac{{p\left( {\mathbf{x}_n^{(t)}|{H_1};\mathbf{0},{\mathbf{\Sigma}_1}} \right)}}{{p\left( {\mathbf{x}_n^{(t)}|{H_0};\mathbf{0},{\mathbf{\Sigma}_0}} \right)}}}\right),
\end{equation}
where
\begin{equation}\label{}
p\left( {\mathbf{x}_n^{(t)}|{H_1};\mathbf{0},{\mathbf{\Sigma} _1}} \right) = \frac{1}{{{\pi ^M}\det ({\mathbf{\Sigma} _1})}}\exp \left( { - (\mathbf{x}_n^{(t)})^H\mathbf{\Sigma} _1^{ - 1}{{\mathbf{x}_n^{(t)}}}} \right)
\end{equation}
and
\begin{equation}\label{}
p\left( {\mathbf{x}_n^{(t)}|{H_0};\mathbf{0},{\mathbf{\Sigma} _0}} \right) = \frac{1}{{{\pi ^M}\det ({\mathbf{\Sigma} _0})}}\exp \left( { - ( \mathbf{x}_n^{(t)}) ^H\mathbf{\Sigma} _0^{ - 1}{{\mathbf{x}_n^{(t)}}}} \right).
\end{equation}

\emph{2) \underline{Optimal LRT with Modulated Ambient Source:}}

When $s_n^{(t)}$ is a discrete modulated signal as defined in (b), we have
\begin{equation}\label{modulated-x(n)}
\mathbf{x}_n^{(t)}\sim
\bigg\{ \begin{matrix}
   \mathcal{C}\mathcal{N}(\mathbf{w}S_q,\sigma _{u}^{2}{\mathbf{I}}_{M}),\;{{H}_{1}} \\
   \mathcal{C}\mathcal{N}(\mathbf{h}S_q,\sigma _{u}^{2}{\mathbf{I}}_{M}),\;{{H}_{0}}   \\
\end{matrix}.
\end{equation}
Similarly, we can then derive the logarithmic form of LRT under modulated ambient source:
\begin{equation}\label{L_M}
L_\mathrm{M}(\mathbf{X}^{(t)}) = \sum\limits_{n = 0}^{N - 1} \ln \left({\frac{\sum\limits_{q = 1}^{Q}{p\left( {\mathbf{x}_n^{(t)}|{H_1};\mathbf{w}S_q,\sigma _{u}^{2}{\mathbf{I}}_{M}} \right)}}{\sum\limits_{q = 1}^{Q}{p\left( {\mathbf{x}_n^{(t)}|{H_0};\mathbf{h}S_q,\sigma _{u}^{2}{\mathbf{I}}_{M}} \right)}}}\right),
\end{equation}
where
\begin{equation}\label{}
p\left( {\mathbf{x}_n^{(t)}|{H_1};\mathbf{w}S_q,{\sigma _{u}^{2}{\mathbf{I}}_{M}}} \right) = \frac{1}{{({\pi}\sigma _{u}^2)^{M}}}\exp \left( { - \frac{1}{\sigma_u^2}(\mathbf{x}_n^{(t)} - \mathbf{w}S_q )^H{{(\mathbf{x}_n^{(t)} - \mathbf{w}S_q)}}} \right)
\end{equation}
and
\begin{equation}\label{}
p\left( {\mathbf{x}_n^{(t)}|{H_0};\mathbf{h}S_q,{\sigma _{u}^{2}{\mathbf{I}}_{M}}} \right) = \frac{1}{{({\pi}\sigma _{u}^2)^{M}}}\exp \left( { - \frac{1}{\sigma_u^2}(\mathbf{x}_n^{(t)} - \mathbf{h}S_q )^H{{(\mathbf{x}_n^{(t)} - \mathbf{h}S_q)}}} \right).
\end{equation}

Based on the above analysis, we can find that although the LRT can achieve the optimal detection performance, it requires the availability of perfect CSI which is not always possible in practical AmBC systems due to the non-cooperation between the legacy transceiver and the reader. On the other hand, the covariance matrix captures various distinguishable features (such as energies and eigenvalues \cite{liu2019deep}) implicitly, it has been adopted to design detectors, as shown in (\ref{L_R}) and (\ref{L_M}). Inspired by this, we propose a tag detection framework based on a DNN which intelligently explores the features of sample covariance matrix.
In the following, we will first introduce the framework and then propose a covariance matrix aware neural network as a realization of the developed framework.

\begin{figure}[t]
  \centering
  \includegraphics[width=0.98\linewidth]{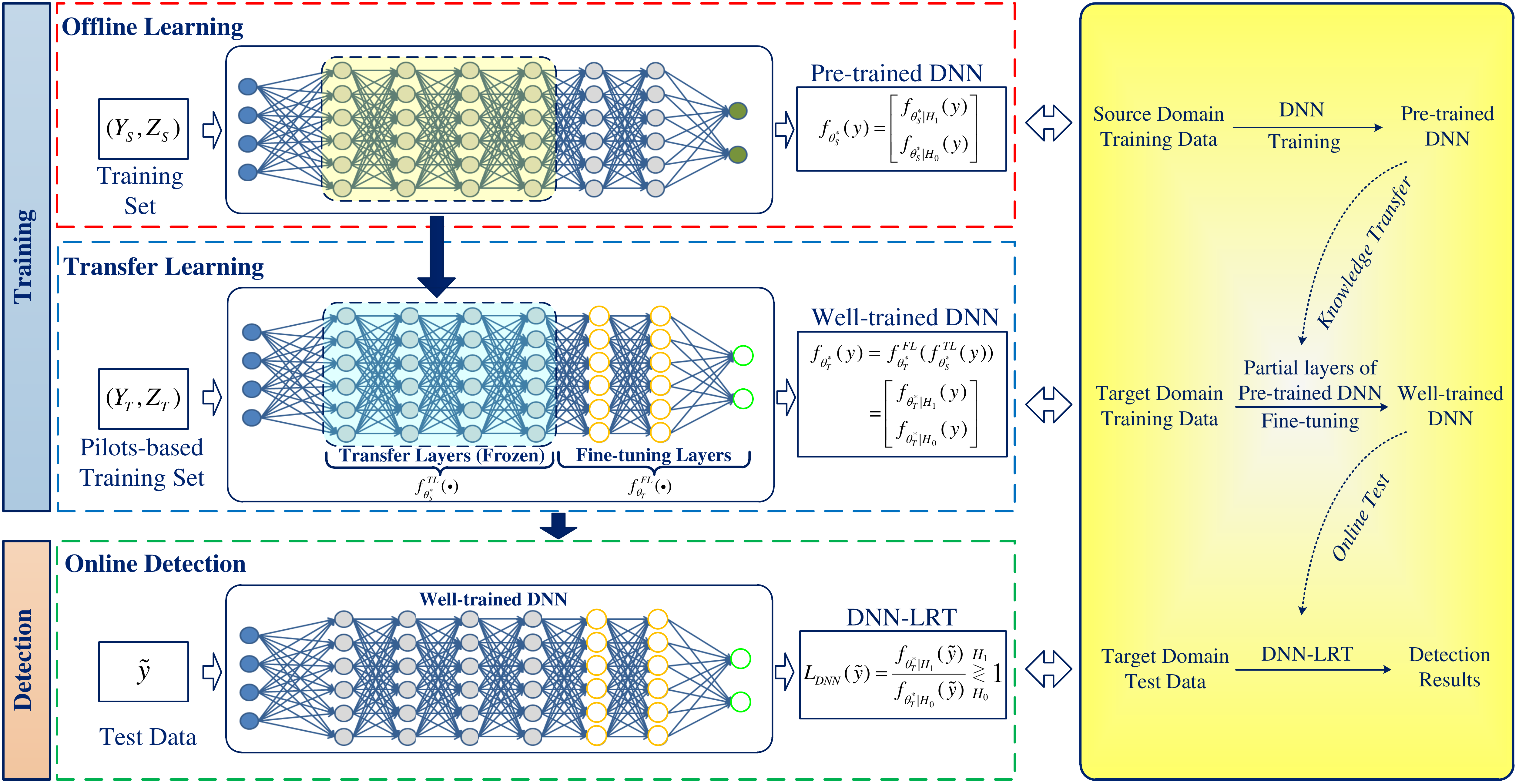}\vspace{-0.6cm}
  \caption{ The proposed DTL framework for tag signal detection. }\vspace{-1cm}
\end{figure}

\section{Deep Transfer Learning-based Tag Signal Detection Framework}
Note that practical channel coefficients change over time due to the time-varying nature of the environment. Based on this, we can adopt a DTL approach to transfer the knowledge obtained from one detection task via offline channel coefficients to another different but related detection task under the real-time channel coefficients. The advanced DTL approach allows the designed detector to adapt itself properly to different channel environments for improving the system performance.
As shown in Fig. 3, we propose a DTL framework, which consists of offline learning, transfer learning, and online detection. {In the proposed framework, we first establish a pre-trained DNN to extract the common features of invariant channel models through offline learning. We then freeze{\footnotemark} \footnotetext{Layer freezing means that the layer weights of a pre-trained neural network keep unchanged during training in a subsequent task, i.e., they remain frozen.} partial layers of the pre-trained DNN and only fine-tune the remaining layers to adjust the DNN to the current channel coefficients through (online) transfer learning. Finally, we can apply the well-trained network for online detection: decoding the tag signals.} {Therefore, the proposed framework does not require the channel information. Instead, it exploits the pilot-based training dataset to capture the features of the real-time channel to facilitate the tag signal detection.} In the following, we will introduce the initialization of deep transfer learning, offline learning module, transfer learning module, and online detection module, respectively.

\subsection{Initialization for Deep Transfer Learning}
The unified dataset adopted in this paper is given by
\begin{equation}\label{unified_dataset}
(Y,Z) = \{({{y}}^{(1)},{{{z}}^{(1)}}),({{y}}^{(2)},{{{z}}^{(2)}}), \cdots ,({{y}}^{(K)},{{{z}}^{(K)}})\},
\end{equation}
where $({{y}}^{(k)},{{{z}}^{(k)}})$ represents the $k$-th, $k\in\{1,2,\cdots,K\}$, example of the training set $(Y,Z)$. In particular, $y^{(k)}$ denotes the input feature which is the test statistic of raw samples and can be regarded as a function of $\mathbf{X}^{(t)}$, as defined in (\ref{observation matrix}). Term $z^{(k)}\in\{1,0\}$ is a label, where ${z}^{(k)}=1$ and ${z}^{(k)}=0$ denote the hypotheses of $H_1$ and $H_0$, respectively.

According to \cite{pan2009survey}, a domain can be defined as $\mathcal{D} = \{\mathcal{Y},P(Y)\}$, where $\mathcal{Y}$ denotes the feature space and $P(Y)$ indicates the marginal probability distribution with $Y = \{ {y}^{(1)}, {y}^{(2)}, \cdots, {y}^{(K)} \} \in \mathcal{Y}$, as defined in (\ref{unified_dataset}).
Correspondingly, the task of $\mathcal{D}$ is defined by $\mathcal{T} = \{\mathcal{Z},P(Z|Y)\}$, where $\mathcal{Z}$ denotes the label space and $P(Z|Y)$ is the posterior probability distribution for $\mathcal{D}$ with $Z = \{ {z}^{(1)}, {z}^{(2)}, \cdots, {z}^{(K)} \} \in \mathcal{Z}$, as defined in (\ref{unified_dataset}). Thus, $P(Z|Y)$ can be regarded as a prediction function $\lambda(\cdot)$ which is used to predict the labels of the inputs.

Note that in this paper, the source domain data and the target domain data arise from the training sets of offline learning and transfer learning, respectively. We can then define the source domain and the target domain as \cite{pan2009survey}
\begin{equation}\label{}
\mathcal{D}_S = \{\mathcal{Y}_S,P(Y_S)\}
\end{equation}
and
\begin{equation}\label{}
\mathcal{D}_T = \{\mathcal{Y}_T,P(Y_T)\},
\end{equation}
respectively.
Here, $\mathcal{Y}_S$ and $\mathcal{Y}_T$ are the feature spaces of the source domain and the target domain, respectively. Terms $P(Y_S)$ and $P(Y_T)$ are the marginal probability distributions of the source domain data and the target domain data, respectively, where $Y_S =\{y_S^{(1)}, y_S^{(2)}, \cdots , y_S^{(K_S)}\} \in \mathcal{Y}_S$ and $Y_T =\{y_T^{(1)}, y_T^{(2)}, \cdots , y_T^{(K_T)}\}$ $\in \mathcal{Y}_T$. Correspondingly, we denote
\begin{equation}\label{}
\mathcal{T}_S = \{\mathcal{Z}_S,P(Z_S|Y_S)\}
\end{equation}
and
\begin{equation}\label{}
\mathcal{T}_T = \{\mathcal{Z}_T,P(Z_T|Y_T)\}
\end{equation}
as the task for $\mathcal{D}_S$ and $\mathcal{D}_T$, respectively. Here, $\mathcal{Z}_S$ and $\mathcal{Z}_T$ denote the label spaces for $\mathcal{D}_S$ and $\mathcal{D}_T$, respectively. Terms $P(Z_S|Y_S)$ and $P(Z_T|Y_T)$ are the posterior probability distributions for $\mathcal{D}_S$ and $\mathcal{D}_T$, respectively, where $Z_S =\{z_S^{(1)}, z_S^{(2)}, \cdots , z_S^{(K_S)}\} \in \mathcal{Z}_S$ and $Z_T =\{z_T^{(1)}, z_T^{(2)}, \cdots , z_T^{(K_T)}\} \in \mathcal{Z}_T$.
Let $\mathbf{\Omega}_{\mathbf{h}_S}$ and $\mathbf{\Omega}_{\mathbf{w}_S}$ (or $\mathbf{\Omega}_{\mathbf{h}_T}$ and $\mathbf{\Omega}_{\mathbf{w}_T}$) denote the sets of channel coefficients $\mathbf{h}$ and $\mathbf{w}$ in the source domain (or the target domain), respectively. In this paper, we assume that the channel coefficients of $\mathcal{D}_S$ and $\mathcal{D}_T$ are i.i.d. with distinct values{\footnotemark}\footnotetext{This i.i.d. assumption is applicable to many practical channel environments. In fact, the proposed framework in this paper is also valid for the case which the channel coefficients of $\mathcal{D}_S$ and $\mathcal{D}_T$ are drawn from different distributions. In this case, it will take more time for the pre-trained network in the source domain to adjust itself to the target domain compared to that of the i.i.d. case.}.
Thus, $\forall i\in [1,|\mathbf{\Omega}_{\mathbf{h}_S}|],j\in [1,|\mathbf{\Omega}_{\mathbf{w}_S}|], p\in [1,|\mathbf{\Omega}_{\mathbf{h}_T}|], q\in [1,|\mathbf{\Omega}_{\mathbf{w}_T}|]$, where $\mathbf{\Omega}(i)$ represents the $i$-th element of the set, we have $\mathbf{\Omega}_{\mathbf{h}_S}(i) \neq \mathbf{\Omega}_{\mathbf{h}_T}(j)$, and $\mathbf{\Omega}_{\mathbf{w}_S}(p) \neq \mathbf{\Omega}_{\mathbf{w}_T}(q)$.
Based on this, it is obvious that $P(Y_S) \neq P(Y_T)$ and we have $\mathcal{D}_S \neq \mathcal{D}_T$.
Therefore, the objective of transfer learning is to improve the learning of the detection problem in $\mathcal{D}_T$ exploiting the knowledge in $\mathcal{D}_S$. To the end, according to \cite{tan2018survey}, we provide the definition of deep transfer learning as follows:

\textbf{Definition 1}. \emph{Deep Transfer Learning (DTL)}. Given a transfer learning task defined by $\langle\mathcal{D}_S, \mathcal{T}_S,$ $\mathcal{D}_T, \mathcal{T}_T, \lambda_T(\cdot)\rangle$. Deep transfer learning aims to improve the performance of predictive function $\lambda_T(\cdot)$ for learning task $\mathcal{T}_T$ by using a deep neural network to discover and transfer latent knowledge from $\mathcal{D}_S$ and $\mathcal{T}_S$, where $\mathcal{D}_S \neq \mathcal{D}_T$ or/and $\mathcal{T}_S \neq \mathcal{T}_T $. \hfill \rule{8pt}{8pt}

According to the definition of DTL and the proposed DTL framework in Fig. 3, the dataset of the offline learning is drawn from the source domain, and the datasets of the transfer learning and the online detection both come from the target domain. In the following, we will systematically introduce the proposed DTL framework based on the definition of DTL.


\subsection{Offline Learning: Pre-trained DNN}
Given the training set of offline learning:
\begin{equation}\label{training_set_expression}
D_S = (Y_S,Z_S) = \{(y_S^{(1)},{z_S^{(1)}}),(y_S^{(2)},{z_S^{(2)}}), \cdots ,(y_S^{(K_S)},{z_S^{(K_S)}})\},
\end{equation}
where $D_S$ denotes the source domain training data, ${y}_S^{(k)}\in \mathcal{D}_S$ and ${{z}_S^{(k)}}\in \mathcal{T}_S$ represent the input and label of the $k$-th, $k\in\{1,2,\cdots,K_S\}$, example, respectively.

Note that the tag detection for AmBC system is actually a binary hypothesis testing problem \cite{qian2017semi}. In this case, the training process becomes a binary classification problem and thus the label is encoded as a one-hot vector \cite{wang2017deep}
\begin{equation}\label{}
{{{\bf{z}}_S}^{(k)}} = \Bigg\{ {\begin{array}{*{20}{c}}
{{{[1,0]}^T},{\kern 1pt} {\kern 1pt} {\kern 1pt} {z_S^{(k)}=1}}\\
{{{[0,1]}^T},{\kern 1pt} {\kern 2pt}  {z_S^{(k)}=0}}
\end{array}}.
\end{equation}
Correspondingly, the output of the DNN is a $2 \times 1$ class score{\footnotemark} vector \cite{goodfellow2016deep}: \footnotetext{{The class score created by a DNN represents the likelihood of each category.}}
\begin{equation}\label{feature_vector}
{f_{\theta_S}}({y}_S^{(k)}) = \left[ {\begin{array}{*{20}{c}}
{{f_{\theta_S|{H_1}}}({y}_S^{(k)})}\\
{{f_{\theta_S|{H_0}}}({y}_S^{(k)})}
\end{array}} \right],
\end{equation}
where ${f_{\theta_S|{H_1}}}({y}_S^{(k)}) + {f_{\theta_S|{H_0}}}({y}_S^{(k)}) = 1$, $0\leq f_{\theta_S}(\cdot) \leq 1$ is the expression of the DNN under the parameter of $\theta_S$, and ${f_{\theta_S|{H_i}}}({y}_S^{(k)})$ represents the $H_i$ class score of the input ${y}_S^{(k)}$ under $f_{\theta_S}(\cdot)$.
In fact, the class score can be rewritten as the following probability expressions \cite{goodfellow2016deep}:
\begin{equation}\label{conditional_pro_single_example}
\begin{array}{l}
{H_1}:{f_{\theta_S|{H_1}}}({y}_S^{(k)}) = P( z_S^{(k)} = 1|{y}_S^{(k)};\theta_S ),\\
{H_0}:{f_{\theta_S|{H_0}}}({y}_S^{(k)}) = P( z_S^{(k)} = 0|{y}_S^{(k)};\theta_S ),
\end{array}
\end{equation}
where $P( z_S^{(k)}|{y}_S^{(k)};\theta_S )$ denotes the conditional probability under $\theta_S$.

Therefore, the goal of the offline learning is to maximize the likelihood
\begin{equation}\label{}
L(\theta_S ) = P(Z_S|Y_S;\theta_S )=\prod\limits_{k = 1}^{K_S} {{{({f_{\theta_S|{H_1}} }({y_S^{(k)}}))}^{{z_S^{(k)}}}}{{( {f_{\theta_S|{H_0}} }({y_S^{(k)}}))}^{1 - {z_S^{(k)}}}}},
\end{equation}
or equivalently the log-likelihood which is given as
\begin{equation}\label{}
l(\theta_S ) = \ln L(\theta_S )= \sum\limits_{k = 1}^{K_S} {z_S^{(k)}}\ln {f_{\theta_S|{H_0}}}({y_S^{(k)}}) + (1 - {z_S^{(k)}})\ln({f_{\theta_S|{H_0}}}({y_S^{(k)}})).
\end{equation}
Therefore, the target of offline learning is to find parameter $\theta_S^*$ to maximize $P(Z_S|Y_S)$, i.e.,
\begin{equation}\label{theta_MAP}
{\theta_S ^*} = \arg {\kern 1pt} {\kern 1pt} \mathop {\max }\limits_{\theta_S}  P( Z_S|Y_S;\theta_S ),
\end{equation}
which is also equivalent to minimizing the following cost function
\begin{equation}\label{cost_function}
J(\theta_S ) =  - \frac{1}{K_S}l(\theta_S )=  - \frac{1}{K_S} \sum\limits_{k = 1}^{K_S} {{z_S^{(k)}}\ln {f_{\theta_S|{H_1}} }({y_S^{(k)}})} + {(1 - {z_S^{(k)}})\ln({f_{\theta_S|{H_0}} }({y_S^{(k)}}))}.
\end{equation}

Based on (\ref{cost_function}), we can then obtain a well-trained DNN through the backpropagation (BP) algorithm \cite{Yi2018Unifying}. In this case, the output of the well-trained DNN can be expressed as
\begin{equation}\label{fv_function}
f_{\theta_S^*}( y ) = \left[ {\begin{array}{*{20}{c}}
{{f_{\theta_S^*|{H_1}}}( y )}\\
{{f_{\theta_S^*|{H_0}}}( y )}
\end{array}} \right],
\end{equation}
where $y$ denotes arbitrary input and $f_{\theta_S^*}( \cdot )$ denotes the well-trained DNN with the well-trained parameter $\theta_S^*$. Correspondingly, ${f_{\theta_S^*|{H_i}}}(y)$ represents the $H_i$ class score of the input $y$ under $f_{\theta_S^*}( \cdot )$.
Since the well-trained DNN is achieved in $\mathcal{D}_S$, we also call it as the pre-trained DNN.
In next subsection, we will introduce transfer learning stage based on the pre-trained DNN.

\subsection{Transfer Learning: Fine-tuning the Pre-trained DNN}
Given the pilots-based set for transfer learning:
\begin{equation}\label{training_set_expression}
D_T  = (Y_T,Z_T) = \{(y_T^{(1)},z_T^{(1)}),(y_T^{(2)},z_T^{(2)}), \cdots ,(y_T^{(K_T)},z_T^{(K_T)})\},
\end{equation}
where $D_T$ denotes the target domain training data{\footnotemark\footnotetext{{Since the pilot symbols and the test data in the same tag frame experience the same channel coefficients in slow fading channels, they have an identical statistical distribution. Thus, the target-domain training data is chosen from the pilot-based data, from which the neural network can learn the features of the test data to facilitate the tag signal detection.}}}, ${y}_T^{(k)}\in \mathcal{D}_T$ and ${z}_T^{(k)}\in \mathcal{T}_T$ represent the input and the label of the $k$-th, $k\in\{1,2,\cdots,K_T\}$, example, respectively. Correspondingly, $y_T^{(k)}$ and $z_T^{(k)}\in\{1,0\}$ are the input of the network and the label, respectively.

Considering both the source task and the target task handle the same tag detection problem, we can reuse the partial feature layers of the pre-trained DNN in the source domain for performance improvement in the target domain, i.e., transferring and transforming partial layers of the pre-trained DNN to be a part of the DNN in the target domain. Based on this, we can then express the DNN in the target domain as
\begin{equation}\label{DNN_target}
  {f_{{\theta _T}}}(y) = f_{{\theta _T}}^{\mathrm{FL}}(f_{\theta _S^*}^{\mathrm{TL}}(y)),
\end{equation}
where ${f_{{\theta _T}}}(\cdot)$ is the target domain's DNN expression with parameter $\theta _T$, $f_{{\theta _T}}^{\mathrm{FL}}(\cdot)$ represents the fine-tuning layers of the target domain's DNN, and $f_{\theta _S^*}^{\mathrm{TL}}(\cdot)$ indicates the transfer layers of the well-trained DNN from the source domain. To achieve (\ref{DNN_target}), we freeze the transfer layers and only update the parameters of the fine-tuning layers during the training process. Similar to (\ref{cost_function}), we define the cost function for the target domain as
\begin{equation}\label{cost_function_target}
J(\theta_T ) =  - \frac{1}{K_T}l(\theta_T ) =  - \frac{1}{K_T} \sum\limits_{k = 1}^{K_T} {{z_T^{(k)}}\ln {f_{\theta_T|{H_0}} }({y_T^{(k)}})} + {(1 - {z_T^{(k)}})\ln(1 - {f_{\theta_T|{H_0}}}({y_T^{(k)}}))}.
\end{equation}
By using the BP algorithm in the training process, we can obtain the well-trained parameter $\theta_T^*$ for the target domain's DNN. Note that $\theta_T^*$ contributes to $\theta_S^*$, hence, the well-trained DNN for the target domain can be expressed as
\begin{equation}\label{fv_function}
f_{\theta_T^*}( y ) = f_{{\theta _T^*}}^{\mathrm{FL}}(f_{\theta _S^*}^{\mathrm{TL}}(y)) = \left[ {\begin{array}{*{20}{c}}
{{f_{\theta_T^*|{H_1}}}( y )}\\
{{f_{\theta_T^*|{H_0}}}( y )}
\end{array}} \right],
\end{equation}
where $f_{\theta_T^*}( \cdot )$ denotes the expression of the well-trained DNN with the well-trained parameter $\theta_T^*$. Function ${f_{\theta_T^*|{H_i}}}(y)$ represents the $H_i$ class score of input $y$ under $f_{\theta_T^*}( \cdot )$.

From a probabilistic viewpoint, if $y \in \mathcal{D}_T$, we can then rewrite the outputs of the well-trained DNN as the following:
\begin{equation}\label{}
\begin{split}
{H_1}: f_{\theta_T^*|{H_1}}(y) = P({H_1}|y), \\
{H_0}: f_{\theta_T^*|{H_0}}(y) = P({H_0}|y), \\
\end{split}
\end{equation}
where $P({H_i}|y)$ denotes the posterior probability expression.
Based on Bayes' theorem, we can obtain the likelihood expressions:
\begin{equation}\label{ConPro_H1}
L({H_1}|y) = \frac{{P({H_1}|y)}\cdot P(y)}{{P({H_1})}} = \frac{{f_{\theta_T^* |{H_1}}^*(y)}\cdot P(y)}{{P({H_1})}}
\end{equation}
and
\begin{equation}\label{ConPro_H0}
L({H_0}|y) = \frac{{P({H_0}|y)}\cdot P(y)}{P({H_0})} = \frac{{f_{\theta_T^* |{H_0}}^*(y)}\cdot P(y)}{{P({H_0})}},
\end{equation}
where $P(H_i)$ is the priori probability of $H_i$, and $P(y)$ is the marginal probability. Note that we always set $P(H_1)=P(H_0)=0.5$ for binary communication systems. Therefore, according to the minimum error probability (MEP) criterion, we can then derive the DTL-based LRT (DTL-LRT):
\begin{equation}\label{T_DNN}
{L}_{\mathrm{DTL}}(y) = \frac{L({H_1}|y)}{L({H_0}|y)} = \frac{f_{\theta_T^*|{H_1}}(y)}{f_{\theta_T^*|{H_0}}(y)} \gtrless 1,
\end{equation}
where we make a decision that $H_1$ holds if ${L}_{\mathrm{DTL}}(y) > 1$, otherwise $H_0$ holds.

\begin{figure}[t]
  \centering
  \includegraphics[width=0.9\linewidth]{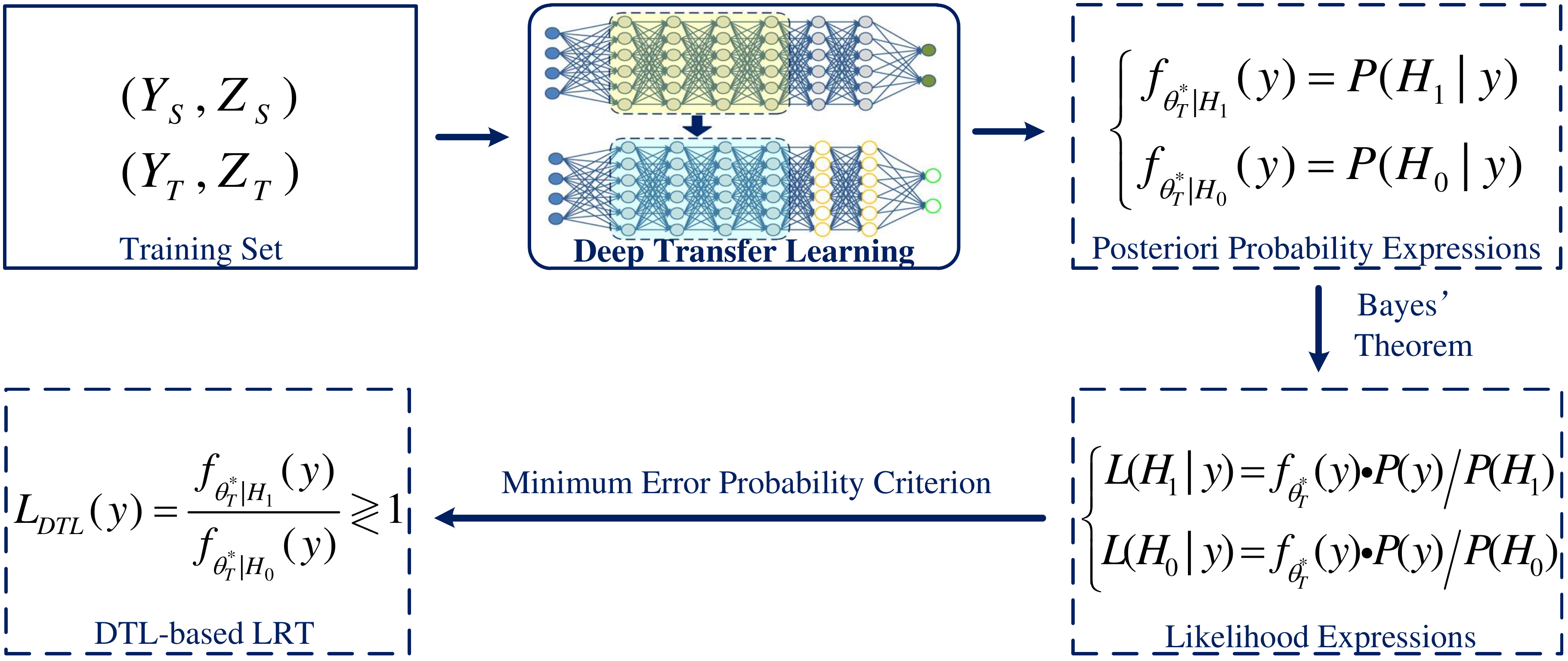}\vspace{-0.6cm}
  \caption{ A flowchart of the derivation of the DTL-based LRT. }\vspace{-1cm}
\end{figure}

For further understanding, we summarize the derivation of DNN-LRT as a flow-chart in Fig. 4. Given the training set, we can first obtain the posterior probability expressions through DTL. Then, based on the Bayes' theorem, we can get the expressions of likelihood of two hypothesis. Finally, according to the minimum error probability (MEP) criterion, we can derive the expression of the DTL-LRT.

\subsection{Online Detection: DTL-LRT using Well-trained DNN }
Given an arbitrary test data $\tilde{y} \in \mathcal{D}_T$, we can directly send it to the well-trained DNN to operate the DTL-LRT, i.e.,
\begin{equation}\label{T_DNN}
{L}_{\mathrm{DTL}}(\tilde{y}) = \frac{f_{\theta_T^*|{H_1}}(\tilde{y})}{f_{\theta_T^*|{H_0}}(\tilde{y})} \gtrless 1,
\end{equation}
where if ${L}_{\mathrm{DTL}}(\tilde{y})>1$, $H_1$ holds, otherwise, $H_0$ holds.

\textbf{Remark 1}: Note that the proposed DTL scheme is a universal DTL workflow for tag signal detection, which is suitable for any neural network and can be realized by any kind of DNN structure, e.g., the multi-layer perceptron, recurrent neural network, and CNN \cite{lecun2015deep}. The proposed DTL scheme does not require to estimate the channel coefficients explicitly. Instead, we only need a few tag signal pilots for transfer learning, which is a practical detection approach for AmBC systems.
{In addition, the proposed method can be extended to the case of $M$-ary tag modulation \cite{van2018ambient, qian2016noncoherent} by extending the current binary classification task to an $M$-class classification task \cite{goodfellow2016deep}, which can further improve the system performance and is left for future work.}

\section{CNN-based Deep Transfer Learning for Tag Signal Detection}
In this section, we provide a realization of the proposed DTL workflow. Note that the sample covariance matrix is a versatile statistic capturing rich distinguishable features. Meanwhile, CNN has a powerful capability in extracting features of matrix-formed data. Therefore, we apply the novel CM-CNN \cite{liu2019deep}, \cite{xie2019activity} to explore the features of the covariance matrix by proposing a covariance matrix-based neural network (CMNet).
In the following, we will introduce the CMNet structure, design the CMNet-based detection algorithm{\footnotemark}{\footnotetext{{Since the backscatter link signal strength from the tag is much weaker than that of the direct link signal from the RF source, the received signals from the tag can be ignored at the legacy receiver \cite{van2018ambient, liu2013ambient, wang2016ambient}.
Thus, the proposed tag signal detection algorithm adopted at the AmBC system does not affect the performance of source signal detection in the legacy system.}}}, and provide the related theoretical analysis.

\begin{figure}[t]
  \centering
  \includegraphics[width=0.96\linewidth]{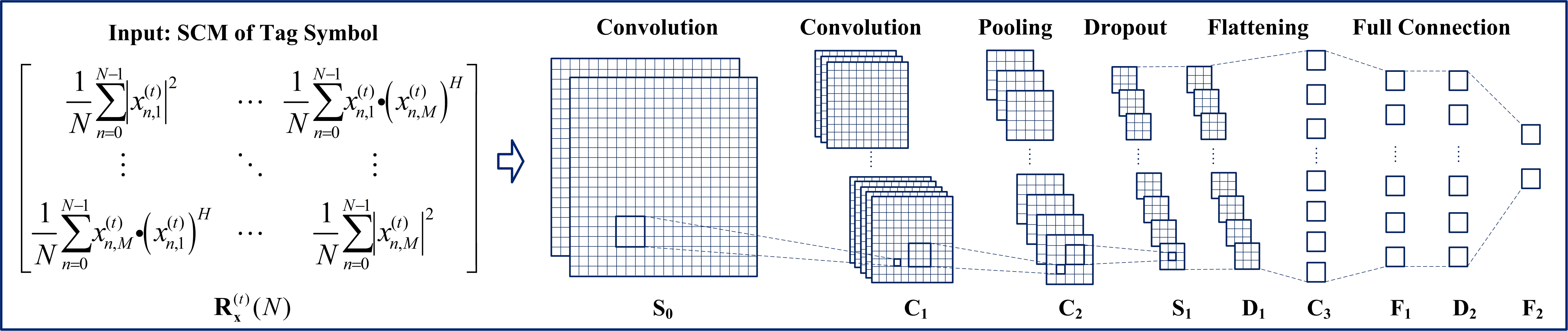}\vspace{-0.6cm}
  \caption{ The designed CMNet structure for tag signal detection. }\vspace{-1cm}
\end{figure}

\subsection{CMNet Structure}
Note that the training set for transfer learning is extracted from pilots with very few samples which may lead to overfitting issue of neural networks \cite{goodfellow2016deep}. To overcome this problem, we introduce two dropout layers and redesign the CM-CNN with a modified structure, as shown in Fig. 5. In particular, CMNet consists of two convolutional layers, one pooling layer, one flattening layer, two dropout layers, and two fully connected layers. The convolutional, pooling, and flattening layers are used for extracting features from input. Then, the dropout layers offer a computationally cheap manner to reduce the possibility of overfitting \cite{srivastava2014dropout}. Finally, the fully connected layers learn the non-linear combinations of these extracted features to further improve the performance of the task.
The corresponding hyperparameters are introduced in Table \Rmnum{1}, where we first set these hyperparameters based on the CM-CNN structure \cite{liu2019deep} and then fine-tune these hyperparameters to search the appropriate values for the network empirically.
The details of each layer will be introduced in the following.

a) \textbf{Input Layer} (${\emph{S}_0}$).

\begin{table}[t]
\normalsize
\caption{Hyperparameters of the proposed CMNet}
\vspace{-0.3cm}
\centering
\small
\renewcommand{\arraystretch}{0.95}
\begin{tabular}{c c}
  \hline
  \vspace{-0.6cm} \\
   \multicolumn{2}{l}{\textbf{Input}: Sample Covariance Matrix (Dimension: $8 \times 8 \times 2$) } \\
  \hline
  \vspace{-0.6cm} \\
  \hspace{0.6cm} \textbf{Layer} &  \hspace{0.6cm} \textbf{Filter Size}   \\
  \vspace{-0.6cm} \\
  \hline
  \vspace{-0.6cm} \\
  \hspace{0.6cm} ${S_0}$ & \hspace{0.6cm}  --   \\
  \vspace{-0.6cm} \\
  \hline
  \vspace{-0.6cm} \\
  \hspace{0.6cm} ${C_1}$ + ReLU & \hspace{0.6cm} $ 32 \times ( 3 \times 3 \times 2 ) $   \\
  \vspace{-0.6cm} \\
  \hline
  \vspace{-0.6cm} \\
  \hspace{0.6cm}  ${C_2}$ + ReLU & \hspace{0.6cm}  $ 32 \times ( 3 \times 3 \times 32 ) $  \\
  \vspace{-0.6cm} \\
  \hline
  \vspace{-0.6cm} \\
  \hspace{0.6cm} $S_1$ (Max-Pooling) & \hspace{0.6cm} $ 2 \times 2 $   \\
  \vspace{-0.6cm} \\

  \hline
  \vspace{-0.6cm} \\
  \hspace{0.6cm} $C_3$ (Flattening) & \hspace{0.6cm} $ 32 \times (4 \times 4) $   \\
  \vspace{-0.6cm} \\
  \hline

  \vspace{-0.6cm} \\
  \hspace{0.6cm} $D_1$ ($\rho = 0.5$) & \hspace{0.6cm} --  \\
  \vspace{-0.6cm} \\
  \hline

  \vspace{-0.6cm} \\
  \hspace{0.6cm} ${F_1}$ + ReLU & \hspace{0.6cm} $ 128 \times 512 $  \\
  \vspace{-0.6cm} \\
  \hline

  \vspace{-0.6cm} \\
  \hspace{0.6cm} $D_2$ ($\rho = 0.25$) & \hspace{0.6cm} --  \\
  \vspace{-0.6cm} \\
  \hline

  \vspace{-0.6cm} \\
  \hspace{0.6cm} ${F_2}$ + Softmax & \hspace{0.6cm} $ 2 \times 128 $  \\
  \vspace{-0.6cm} \\
  \hline
  \vspace{-0.6cm} \\
   \multicolumn{2}{l}{\textbf{Output}: Score Vector (Dimension: $2 \times 1$)} \\
  \vspace{-0.6cm} \\
  \hline
\end{tabular}
\vspace{-0.6cm}
\end{table}

According to (\ref{observation matrix}), there are $N$ RF source signal sampling periods within one tag symbol, it is naturally to compute the sample covariance matrix of the $t$-th tag symbol as:
\begin{equation}\label{Rx(N)}
{{\bf{R}}_{\bf{x}}^{(t)}}(N) = \frac{1}{N}{\mathbf{X}^{(t)}}{(\mathbf{X}^{(t)})}^{H}=\frac{1}{N}\sum\limits_{n = 0}^{N - 1} {{\bf{x}}_n^{(t)}{({\bf{x}}_n^{(t)})^H}}.
\end{equation}
Considering ${{\bf{R}}_{\bf{x}}^{(t)}}(N)$ is a complex-valued matrix, it is necessary to adopt two input channels to represent the real part and imaginary part, respectively. Let $S_0(i; j; \beta) \in \mathbb{R}^{M \times M}$ represent the $i$-th row and the $j$-th column element of the $\beta$-th channel in $S_0$ layer, we have
\begin{equation}\label{model_S0}
S_0(i; j; 0) = (\mathrm{Re}({{\bf{R}}_{\bf{x}}^{(t)}}(N))_{i,j}
\end{equation}
and
\begin{equation}\label{}
S_0(i; j; 1) = (\mathrm{Im}({{\bf{R}}_{\bf{x}}^{(t)}}(N))_{i,j}.
\end{equation}

b) \textbf{Convolutional Layers} (${{\emph{C}_1}}$,${{\emph{C}_2}}$, and ${{\emph{C}_3}}$).

We adopt three convolutional layers, denoted by ${{\emph{C}_1}}$,${{\emph{C}_2}}$, and ${{\emph{C}_3}}$, to further explore the features for detection. Each feature map is the convolution results of the last layer. As shown in Table \Rmnum{1}, $p \times ( q \times q )$ denotes that there are $p$ kernels with kernel size of $ q \times q $.
Take $C_1$ as an example, if we adopt a convolution size of $ L \times L $, the $i$-th row and the $j$-th column element of the $\beta$-th feature map in $C_1$ can be expressed as
{\begin{equation}\label{model_C1}
{C_1}{\rm{(}}i,j,\beta {\rm{)}} = {f_R}\big( \sum\limits_{r = 0}^{1} \sum\limits_{{i_0} = 0}^{L-1} \sum\limits_{{j_0} = 0}^{L-1} [{S_0}(i + {i_0},j + {j_0},r) \cdot {K_\beta^{C_1}}(L - {i_0},L - {j_0},1 - r)]    \big), \\
\end{equation}
where $f_R(t)=\max(0,t)$ represents the ReLU function and ${K_\beta^{C_1}}(\cdot,\cdot,\cdot)$ indicates the corresponding kernel of the $\beta$-th feature map in $C_1$. In Table \Rmnum{1}, we adopt ``ReLU'' and ``Softmax'' to denote the activation functions using rectified linear unit and normalized exponential function \cite{goodfellow2016deep}, respectively. In addition, ``flattening'' means that we flatten the outputs of the convolutional layer to create a single long feature vector for the following fully connected layers.

c) \textbf{Pooling Layer} (${{\emph{S}_1}}$).

There is one pooling layer, i.e., ${{\emph{S}_1}}$, which is obtained by the maximum polling operations of the last convolutional layer. As shown in Table \Rmnum{1}, the pooling size of $S_1$ is $2 \times 2$, and ``Max-Pooling'' refers to the the maximum polling operation.
Based on above analysis, the $i$-th row and the $j$-th column element of the $\beta$-th feature map in $S_1$ can be expressed as
\begin{equation}\label{S1_pooling}
S_1(i;j;\beta) = \max(C_1(2i-1,2j-1,\beta), C_1(2i-1,2j,\beta),C_1(2i,2j-1,\beta), C_1(2i,2j,\beta)).
\end{equation}

d) \textbf{Dropout Layers} (${{\emph{D}_1}}$ and ${{\emph{D}_2}}$).
Considering that there are limited training examples for transfer learning, we introduce two dropout layers $D_1$ and $D_2$ to overcome the overfitting problem. In particular, the dropout rate, denoted by $\rho$ in Table \Rmnum{1}, is a common dropout hyperparameter, which refers to the probability of training a given node in a layer. For example, $\rho = 0$ indicates no dropout and $\rho = 1.0$ means that there are no outputs from the layer.

e) \textbf{Fully Connected Layers} (${{\emph{F}_1}}$ and ${{\emph{F}_2}}$).
To fully exploit of the feature maps, one fully connected layer $F_1$ is connected with flattening layer $C_3$, and the number of neurons in $F_1$ is set as $128$. Finally, a two-neuron fully connected layer $F_2$ with a softmax function is connected as the output of CMNet. Therefore, we can regard the network as a non-linear function and express the CMNet as
\begin{equation}\label{expression_CMNet}
{h}_{\theta_S}( {{\bf{R}}_{\bf{x}}^{(t)}}(N) ) = \left[ {\begin{array}{*{20}{c}}
{{{h}_{\theta_S |{H_1}}}( {{\bf{R}}_{\bf{x}}^{(t)}}(N) )}\\
{{{h}_{\theta_S |{H_0}}}( {{\bf{R}}_{\bf{x}}^{(t)}}(N) )}
\end{array}} \right],
\end{equation}
where ${{h}_{\theta_S}( \cdot )}$ is the expression of the CMNet and ${{h}_{\theta_S | {H_i}}( \cdot )}$ is the class score of $H_i$ by the CMNet.

\textbf{Remark 2}: Note that although the dimension of the sample covariance matrix generally changes with the number of antennas, the high scalability of the proposed CMNet structure can be easily extended to different shapes accordingly, as will be verified in the simulation section.

\begin{figure}[t]
  \centering
  \includegraphics[width=0.96\linewidth]{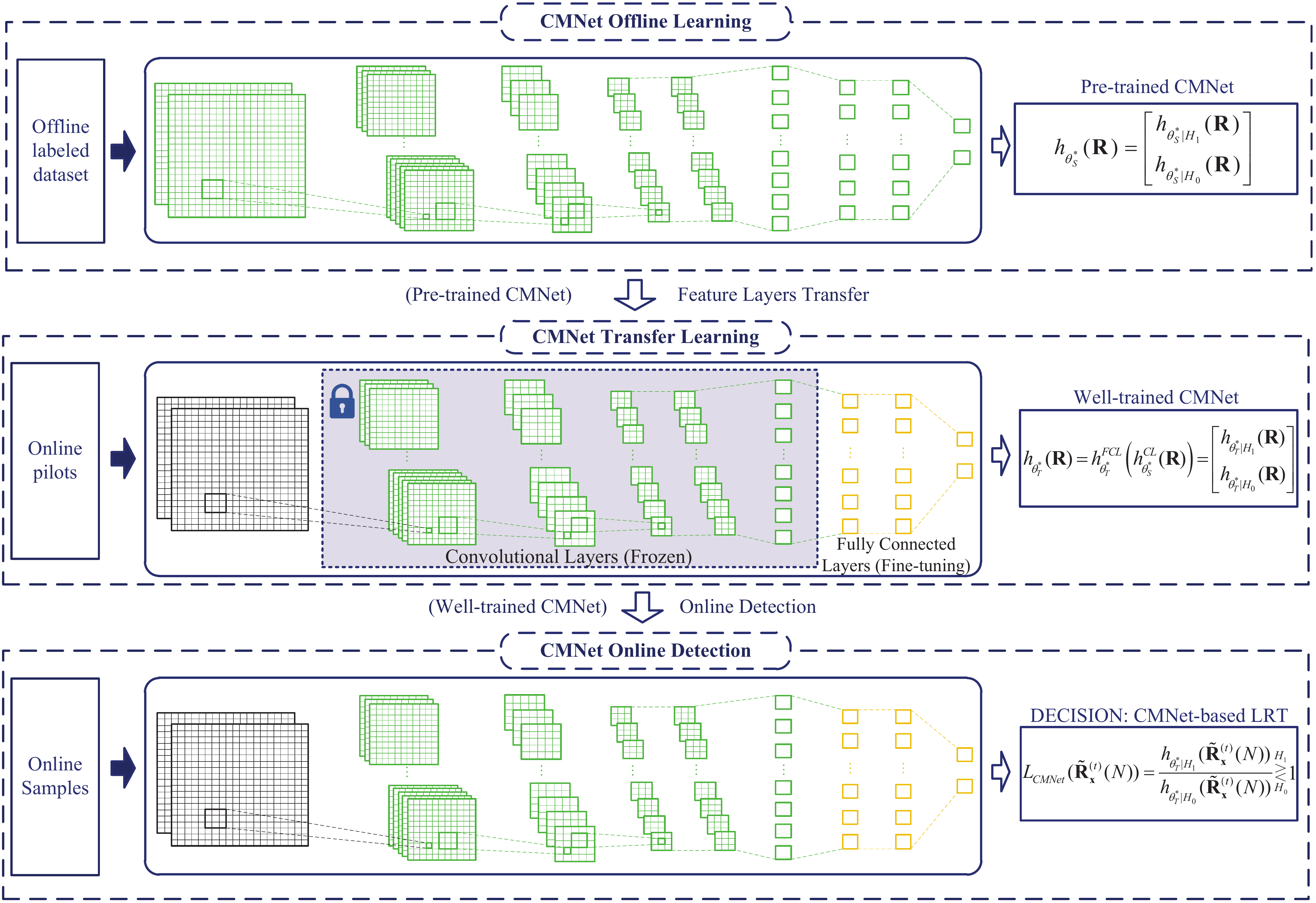}\vspace{-0.6cm}
  \caption{ The proposed CMNet-based deep transfer learning scheme for tag signal detection (A realization of the developed DTL-based workflow in Fig. 3). }\vspace{-1cm}
\end{figure}

\subsection{ CMNet-based Detection Algorithm }
In this section, we adopt the designed CMNet as a core DNN of the developed framework in Fig. 3 to perform tag detection. The details are summarized in Fig. 6. To start with, we first perform offline learning.

\subsubsection{Offline Learning}
Given $K_S$ labelled tag symbols{\footnotemark}\footnotetext{{Historically, researchers
have designed many effective channel models which can well characterize the actual channels in terms of channel statistics. With these well-developed channel models, the training data of offline learning can be obtained by simulation \cite{ye2017power}.}}, we adopt $\mathbf{X}_S^{(k)}=[{\mathbf{x}^{(k)}_{S_1}},{\mathbf{x}^{(k)}_{S_2}},\cdots,{\mathbf{x}^{(k)}_{S_N}}] \in \mathbb{C}^{M\times N}$ to denote the sampling matrix of the $k$-th, $k\in\{1,2,\cdots,K_S\}$, tag symbol in the source domain. According to (\ref{Rx(N)}), the sample covariance matrix of the $k$-th tag symbol is written as
\begin{equation}\label{R_S(N)}
{{\bf{R}}_{{\bf{x}}_S}^{(t)}}(N) = \frac{1}{N}{\mathbf{X}_S^{(t)}}{(\mathbf{X}_S^{(t)})}^{H}.
\end{equation}
Based on this, we build the training set of the source domain:
\begin{equation}\label{}
D_S =(\mathbf{\Omega }_S,Z_S)=\{({\bf{R}}_{\bf{x}}^{(1)}(N),{{{z}}_S^{(1)}}),({\bf{R}}_{\bf{x}}^{(2)}(N),{{{z}}_S^{(2)}}), \cdots ,({\bf{R}}_{\bf{x}}^{(K_S)}(N),{{{z}}_S^{(K_S)}})\},
\end{equation}
where ${{{z}_S}^{(k)}} \in \{0,1\}$ is the corresponding label.

According to (\ref{cost_function}), we can then derive the cost function of CMNet as
\begin{equation}\label{CMCNNv2_cost_function_Source}
{J_{\mathrm{CMNet}}}(\theta_S ) =  - \frac{1}{K_S}\sum\limits_{k = 1}^{K_S} {{z_S^{(k)}}\ln ({{h}_{\theta_S|{H_1} }}({{{\bf{R}}_{{\bf{x}}_S}^{(k)}(N)}}))} + (1 - {z_S^{(k)}})\ln({{h}_{\theta_S|{H_0}} }({{\bf{R}}_{{\bf{x}}_S}^{(k)}(N)})).
\end{equation}
Then, exploiting the BP algorithm, we can obtain the pre-trained CMNet:
\begin{equation}\label{fv_CMNet}
{h}_{\theta_S^*}( \mathbf{R} ) = \left[ {\begin{array}{*{20}{c}}
{{{h}_{\theta_S^* |{H_1}}}( \mathbf{R} )}\\
{{{h}_{\theta_S^* |{H_0}}}( \mathbf{R} )}
\end{array}} \right],
\end{equation}
where $\mathbf{R}$ is the input matrix which can be an arbitrary sample covariance matrix and ${\hat{h}_{\theta_S^*}( \cdot )}$ is the expression of the pre-trained CMNet.

\subsubsection{Transfer Learning}
According to the frame structure in Fig. 2, there are $P$ pilots for transfer learning. For any pilot, we use $\mathbf{X}_T^{(k)}=[{\mathbf{x}^{(k)}_{T_1}},{\mathbf{x}^{(k)}_{T_2}},\cdots,{\mathbf{x}^{(k)}_{T_N}}]
$ to denote the target domain's sampling matrix of the $k$-th, $k\in\{1,2,\cdots,K_T\}$ tag symbol. Similar to (\ref{Rx(N)}), we have
\begin{equation}\label{R_S(N)}
{{\bf{R}}_{{\bf{x}}_T}^{(k)}}(N) = \frac{1}{N}{\mathbf{X}_T^{(k)}}{(\mathbf{X}_T^{(k)})}^{H}.
\end{equation}
We can then build the training set of the target domain:
\begin{equation}\label{}
D_T =(\mathbf{\Omega }_T,Z_T)=\{({\bf{R}}_{{\bf{x}}_T}^{(1)}(N),{{{z}}_T^{(1)}}),({\bf{R}}_{{\bf{x}}_T}^{(2)}(N),{{{z}}_T^{(2)}}), \cdots ,({\bf{R}}_{{\bf{x}}_T}^{(K_T)}(N),{{{z}}_T^{(K_T)}})\},
\end{equation}
where ${{z}_T^{(k)}} \in \{0,1\}$ is the corresponding label.

Similar to (\ref{cost_function}), we define the CMNet's cost function for target domain as
\begin{equation}\label{CMCNNv2_cost_function_Target}
{J_{\mathrm{CMNet}}}(\theta_T ) =  - \frac{1}{K_T}\sum\limits_{k = 1}^{K_T} {{z_T^{(k)}}\log ({{h}_{\theta_T|{H_1} }}({{{\bf{R}}_{{\bf{x}}_T}^{(k)}(N)}}))} + (1 - {z_T^{(k)}})\log({{h}_{\theta_T|{H_0}} }({{\bf{R}}_{{\bf{x}}_T}^{(k)}(N)})).
\end{equation}
Based on the cost function, we can then operate the training process. As analyzed in (\ref{DNN_target}), we need to freeze the convolutional layers and only update the parameters of the fully connected layers by using the backpropagation algorithm, as analyzed in Section \Rmnum{3}. Finally, we can obtain the well-trained CMNet:
\begin{equation}\label{fv_function}
h_{\theta_T^*}( \mathbf{R} ) = h_{{\theta _T^*}}^{\mathrm{FCL}}(h_{\theta _S^*}^{\mathrm{CL}}(\mathbf{R})) = \left[ {\begin{array}{*{20}{c}}
{{h_{\theta_T^*|{H_1}}}( \mathbf{R} )}\\
{{h_{\theta_T^*|{H_0}}}( \mathbf{R} )}
\end{array}} \right],
\end{equation}
where $h_{\theta_T^*}( \cdot )$ denotes the expression of the well-trained CMNet with the well-trained parameter $\theta_T^*$, $h_{{\theta _T^*}}^{\mathrm{FCL}}(\cdot)$ denotes the fully connected layers ($F_1\rightarrow F_2$), and $f_{\theta _S^*}^{\mathrm{CL}}(\cdot)$ represents the convolutional layers ($C_1\rightarrow C_3$).

\subsubsection{Online Detection}
Given the $t$-th tag symbol's sampling matrix for testing, denoted by $\mathbf{\tilde{X}}^{(t)}\in\mathbb{C}^{M \times N}$, we can obtain the corresponding sample covariance matrix $\mathbf{\tilde{R}_x}^{(t)}(N) = \frac{1}{N}\mathbf{\tilde{X}}^{(t)}(\mathbf{\tilde{X}}^{(t)})^{H}$. From the well-trained CMNet, we then derive the CMNet-based LRT as
\begin{equation}\label{L-CMNet}
{L}_{\mathrm{CMNet}}(\mathbf{\tilde{R}_x}^{(t)}(N)) = \frac{h_{\theta_T^*|{H_1}}(\mathbf{\tilde{R}_x}^{(t)}(N))}{h_{\theta_T^*|{H_0}}(\mathbf{\tilde{R}_x}^{(t)}(N))} \mathop \gtrless\limits_{{c^{(t)}=0}}^{{c^{(t)}=1}} 1,
\end{equation}
where we make a decision that $c^{(t)}=1$ if ${L}_{\mathrm{CMNet}}(\mathbf{\tilde{R}_x}^{(t)}(N))>1$, otherwise, $c^{(t)}=0$.

\subsubsection{Algorithm Steps}
Based on the analysis above, we propose a novel CMNet-based detection algorithm, which is summarized in \textbf{Algorithm 1}, where $i_S$ and $i_T$ are iteration indices, and $I_S$ and $I_T$ denote the maximum numbers of iterations of the source domain and the target domain, respectively.


\begin{table}[t]
\normalsize
\vspace{-0.3cm}
\centering
\renewcommand{\arraystretch}{1.0}
\begin{tabular}{l}
\vspace{-0.35cm}\\
\toprule[1.8pt] \vspace{-0.85cm}\\
\hspace{-0.1cm} \textbf{Algorithm 1} \hspace{0.6cm} {CMNet-based Detection Algorithm} \hspace{0.1cm} \\
\toprule[1.8pt] \vspace{-0.85cm}\\
\textbf{Initialization:} \\
\hspace{1.8cm} $i_S = 0$, $i_T = 0$, $\theta_S$ with random weights, $\theta_T$ with random weights   \\
\textbf{Offline Learning:} \\
1:\hspace{1.35cm}\textbf{Input:} Training set $D_S =(\mathbf{\Omega }_S,Z_S)$\\
2:\hspace{1.7cm}\textbf{while} $i_S \leq I_S $ \textbf{do} \\
3:\hspace{2.2cm}update $\theta_S$ by BP algorithm on $J_{\mathrm{CMNet}}(\theta_S)$ in (\ref{CMCNNv2_cost_function_Source})\\
\hspace{2.4cm} $i_S = i_S + 1$  \\
4:\hspace{1.7cm}\textbf{end while} \\
5:\hspace{1.35cm}\textbf{Output}:  ${h}_{\theta_S^*}( \cdot ) $\\
\textbf{Transfer Learning:} \\
6:\hspace{1.35cm}\textbf{Input:} Training set $D_T =(\mathbf{\Omega }_T,Z_T)$\\
7:\hspace{1.7cm}\textbf{while} $i_T \leq I_T $ \textbf{do} \\
8:\hspace{2.2cm}update $\theta_T$ by BP algorithm on $J_{\mathrm{CMNet}}(\theta_T)$ in (\ref{CMCNNv2_cost_function_Target})\\
\hspace{2.4cm} $i_T = i_T + 1$  \\
9:\hspace{1.7cm}\textbf{end while} \\
10:\hspace{1.2cm}\textbf{Output:}  ${h}_{\theta_T^*}( \cdot ) $\\
\textbf{Online Detection:} \\
11:\hspace{1.2cm}\textbf{Input:} Test data $\mathbf{\tilde{R}}_x$ \\
12:\hspace{1.55cm}\textbf{do} CMNet-LRT by (\ref{L-CMNet}) \\
13:\hspace{1.2cm}\textbf{Output:} Decision: $c^{(t)}=1$ or $c^{(t)}=0$ \\
\bottomrule[1.8pt]
\end{tabular}\vspace{-0.6cm}
\end{table}

\subsection{Theoretical Analysis}
{In general, a practical neural network consists of a large number of non-linear units and parameters, which is intractable for analysis \cite{goodfellow2016deep, wang2017deep, mao2018deep}.
As an alternative, we adopt an indirect approach to analyze the performance of the proposed approach.
In particular, to shed light on the performance of the considered system, in this paper, we formulate the proposed CMNet as a non-linear function and analyze the output of CMNet asymptotically to characterize its properties when the number of samples is sufficiently large under a richly scattered multipath environment{\footnotemark}\footnotetext{This assumption is merely made for convenience of the following analysis and the proposed method is valid for any channel models satisfying the previously stated conditions.}.}
In this case, the received sampling vector at the multi-antenna reader can be formulated as \cite{tse2005fundamentals}
\begin{equation}\label{}
\begin{split}
 & {{H}_{1}}:\mathbf{x}_n^{(t)} = \tilde{{\mathbf{s}}}_{\mathbf{w}n}^{(t)} + \mathbf{u}_n^{(t)}, \\
 & {{H}_{0}}:\mathbf{x}_n^{(t)} = \tilde{{\mathbf{s}}}_{\mathbf{h}n}^{(t)} + \mathbf{u}_n^{(t)}, \\
\end{split}
\end{equation}
where $\tilde{{\mathbf{s}}}_{\mathbf{w}n}^{(t)} = \mathbf{w}s_n^{(t)} \in \mathbb{C}^{M\times1}$ (or $\tilde{{\mathbf{s}}}_{\mathbf{h}n}^{(t)} = \mathbf{h}s_n^{(t)} \in \mathbb{C}^{M\times1}$) can be modelled by i.i.d. Gaussian random vector with $\mathcal{CN} (\mathbf{0}, \sigma_{sw}^2\mathbf{I}_M)$ (or $\mathcal{CN} (\mathbf{0}, \sigma_{sh}^2\mathbf{I}_M)$). Based on this, the distribution of $\mathbf{x}_n^{(t)}$ can be expressed by
\begin{equation}\label{}
\mathbf{x}_n^{(t)}\sim
\bigg\{ \begin{matrix}
   \mathcal{C}\mathcal{N}(\mathbf{0},{\sigma_1^2{{\bf{I}}_M}}),{{H}_{1}}, \\
   \mathcal{C}\mathcal{N}(\mathbf{0},{\sigma_0^2{{\bf{I}}_M}}),{{H}_{0}}, \\
\end{matrix}
\end{equation}
where $\sigma_1^2 = \sigma_{sw}^2 + \sigma_u^2$ and $\sigma_0^2 = \sigma_{sh}^2 + \sigma_u^2$. In this case, if we ignore the irrelevant items in (\ref{L_R}), the optimal LRT can be expressed as
\begin{equation}\label{T-EN}
  T_{\mathrm{LRT}} = \sum\limits_{n = 0}^{N - 1} {{{\left\| {{\bf{x}}(n)} \right\|}^2}}.
\end{equation}
Therefore, when considering a richly scattered environment, the optimal LRT detector behaves the same as an energy detector (ED) with perfect CSI.

In the following, we will analyze the output of the CMNet.
When the number of samples is sufficiently large, the sample covariance matrix has the following expression:
\begin{equation}\label{Sigma_IM}
{{\mathbf{R}}_{\mathbf{x}}^{(t)}}(N) \mathop \approx \limits^{N \rightarrow \infty } \bigg\{ {\begin{array}{*{20}{l}}
{\sigma_1^2{{\bf{I}}_M},\hspace{0.2cm}{H_1}}, \\
{\sigma_0^2{{\bf{I}}_M}, \hspace{0.2cm}{H_0}}.
\end{array}}
\end{equation}
Hence, the sample covariance matrix approaches a real-valued diagonal matrix, that is, the real part and imaginary part of ${{\mathbf{R}}_{\mathbf{x}}^{(t)}}(N)$ become a diagonal matrix and a zero matrix, respectively. According to (\ref{model_S0}), we can express the element of input layer $S_0$ as \vspace{-0.1cm}
\begin{equation}\label{}
{S_0}(i,j,0){\rm{ = }}\bigg\{ {\begin{array}{*{20}{l}}
{{\sigma ^2},i = j},\\
{0\;\;,i \neq j},
\end{array}}
\end{equation}
where $\sigma^2$ can be either $\sigma_1^2$ or $\sigma_0^2$. Similar to (\ref{model_C1}), if we select a kernel size of $L \times L$, the element at the $i$-th row and the $j$-th column of the $\beta$-th feature map in $C_1$ can be expressed as \vspace{-0.1cm}
\begin{equation}\label{C1_sigma}
\begin{array}{l}
{C_1}{\rm{(}}i,j,\beta {\rm{)}} = {f_R}\left( {\sum\limits_{{i_0} = 0}^{L-1} {\sum\limits_{{j_0} = 0}^{L-1} {[{S_0}(i + {i_0},j + {j_0},0) \cdot {K_\beta^{C_1} }(L - {i_0},L - {j_0},0)]} } } \right) \vspace{0.2cm}\\
{\kern 9pt} {\kern 1pt} {\kern 1pt} {\kern 1pt} {\kern 1pt} {\kern 1pt} {\kern 1pt} {\kern 1pt} {\kern 1pt} {\kern 1pt} {\kern 1pt} {\kern 1pt} {\kern 1pt} {\kern 1pt} {\kern 1pt} {\kern 1pt} {\kern 1pt} {\kern 1pt} {\kern 1pt} {\kern 1pt} {\kern 1pt} {\kern 1pt} {\kern 1pt} {\kern 1pt} {\kern 1pt} {\kern 1pt} {\kern 1pt} {\kern 1pt} {\kern 1pt} {\kern 1pt} {\kern 1pt} {\kern 1pt} {\kern 1pt} {\kern 1pt} {\kern 1pt} {\kern 1pt} {\kern 1pt} {\kern 1pt} {\kern 1pt} {\kern 1pt} {\kern 1pt} {\kern 1pt} {\kern 1pt} {\kern 1pt} {\rm{ = }}{f_R}\left( {\sum\limits_{d = 1}^M {\sum\limits_{\scriptstyle i + {i_0} = j + {j_0} = d,\hfill\atop
\scriptstyle0 \le {i_0},{j_0} \le L-1\hfill} {[{S_0}(d,d,0) \cdot {K_\beta^{C_1} }(L - {i_0},L - {j_0},0)]} } } \right)\vspace{0.2cm}\\
{\kern 9pt} {\kern 1pt} {\kern 1pt} {\kern 1pt} {\kern 1pt} {\kern 1pt} {\kern 1pt} {\kern 1pt} {\kern 1pt} {\kern 1pt} {\kern 1pt} {\kern 1pt} {\kern 1pt} {\kern 1pt} {\kern 1pt} {\kern 1pt} {\kern 1pt} {\kern 1pt} {\kern 1pt} {\kern 1pt} {\kern 1pt} {\kern 1pt} {\kern 1pt} {\kern 1pt} {\kern 1pt} {\kern 1pt} {\kern 1pt} {\kern 1pt} {\kern 1pt} {\kern 1pt} {\kern 1pt} {\kern 1pt} {\kern 1pt} {\kern 1pt} {\kern 1pt} {\kern 1pt} {\kern 1pt} {\kern 1pt} {\kern 1pt} {\kern 1pt} {\kern 1pt} {\kern 1pt} {\kern 1pt} {\kern 1pt} {\rm{ = }}{\sigma ^2}{f_R}\left( {\sum\limits_{d = 1}^M {\sum\limits_{\scriptstyle i + {i_0} = j + {j_0} = d,\hfill\atop
\scriptstyle0 \le {i_0},{j_0} \le L-1\hfill} {{K_\beta^{C_1} }(L - {i_0},L - {j_0},0)} } } \right).
\end{array}
\end{equation}
For a well-trained CMNet, the parameters are fixed and thus (\ref{C1_sigma}) can be rewritten as \vspace{-0.2cm}
\begin{equation}\label{}
{C_1}(i,j,\beta) = \eta \sigma ^2,
\end{equation}
where
\begin{equation}\label{}
\eta = f_R \left( {\sum\limits_{d = 1}^M {\sum\limits_{\scriptstyle i + {i_0} = j + {j_0} = d,\hfill\atop
\scriptstyle0 \le {i_0},{j_0} \le L-1\hfill} {{K_\beta^{C_1} }(L - {i_0},L - {j_0},0)} } } \right)
\end{equation}
is a constant term.
Therefore, if we use ${h}_{\theta_T^*}^{C_1}(\cdot)$ to represent the expression of $C_1$ of the well-trained CMNet, we have the homogeneity property: ${h}_{\theta_T^*}^{C_1} ({\sigma}^2\mathbf{I}_M) = {\sigma}^2 {h}_{\theta_T^*}^{C_1} (\mathbf{I}_M)$.

Based on the above discussions, if we consider $C_1$ to $F_1$ as a subnetwork ${h}_{\theta_T^*}^{{C_1}{F_1}} (\cdot)$, we have
\begin{equation}\label{}
{h}_{\theta_T^*}^{{C_1}{F_1}}({\sigma ^2}{{\mathbf{I}}_M})= {\sigma ^2}{h}_{\theta_T^*}^{{C_1}{F_1}}({{\mathbf{I}}_M}).
\end{equation}
Denote $\tilde{\theta} = \{\tilde{\theta}_1,\tilde{\theta}_2\}$ as the weights between $F_1$ and $F_2$ of the well-trained CMNet, where $\tilde{\theta}_1$ and $\tilde{\theta}_2$ indicate the weights corresponding to the the 1-th row and 2-th row elements of $F_2$, respectively. The output of the well-trained CMNet can be finally expressed as
\begin{equation}\label{h_theo}
\begin{split}
{{h}_{\theta_T^*} }({\sigma ^2}{{\mathbf{I}}_M}) & = {f_{\mathrm{softmax}}}\left( {{\sigma ^2}{{\tilde \theta }^{T}}{h}_{\theta_T^*}^ {{C_1}{F_1}}({{\mathbf{I}}_M})} \right)\\
& =
\frac{1}{ \exp({\sigma ^2} \tilde{\theta}_1^{T}{{h}_{\theta_T^*}^ {{C_1}{F_1}}({{\mathbf{I}}_M})})  + \exp( {\sigma ^2} \tilde{\theta}_2^{T} {
{h}_{\theta_T^*}^ {{C_1}{F_1}}({{\mathbf{I}}_M})} ) }
\begin{bmatrix}
\exp( {\sigma ^2} \tilde{\theta}_1^{T} {{h}_{\theta_T^*}^ {{C_1}{F_1}}({{\mathbf{I}}_M})} ) \\
\exp( {\sigma ^2} \tilde{\theta}_2^{T} {{h}_{\theta_T^*}^ {{C_1}{F_1}}({{\mathbf{I}}_M})} ) \\
\end{bmatrix},
\end{split}
\end{equation}
where
\begin{equation}\label{}
\begin{split}
f_{\mathrm{softmax}}(x) &=
\frac{1}{ \exp(\tilde{\theta}_1^{T}x)  + \exp( \tilde{\theta}_2^{T} x ) }
\begin{bmatrix}
\exp( \tilde{\theta}_1^{T} x ) \\
\exp( \tilde{\theta}_2^{T} x )
\end{bmatrix}
\end{split}
\end{equation}
denotes the softmax function.

According to (\ref{L-CMNet}), the CMNet-based LRT can be expressed as
\begin{equation}\label{L-CMNet-theo}
  L_{\mathrm{CMNet}} = \frac{\exp( {\sigma ^2} \tilde{\theta}_1^{T} {{h}_{\theta_T^*}^ {{C_1}{F_1}}({{\mathbf{I}}_M})} )}{\exp( {\sigma ^2} \tilde{\theta}_2^{T} {{h}_{\theta_T^*}^ {{C_1}{F_1}}({{\mathbf{I}}_M})} )}= \exp({\sigma^2(\tilde{\theta}_1^{T} - \tilde{\theta}_2^{T}) {h}_{\theta_T^*}^ {{C_1}{F_1}} (\mathbf{I}_M)
  }) \gtrless 1.
\end{equation}
Note that when $N$ is large enough, $\sigma^2 \approx \sum\limits_{n = 0}^{N - 1} {{{\left\| {{\bf{x}}(n)} \right\|}^2}} / {MN}$ \cite{kay1998fundamentals}.
Discard the irrelevant constants, we can rewrite (\ref{L-CMNet-theo}) as
\begin{equation}\label{L-CMNet-EN}
  \sum\limits_{n = 0}^{N - 1} {{{\left\| {{\bf{x}}(n)} \right\|}^2}} \gtrless \gamma,
\end{equation}
where $\gamma$ is the threshold. {Therefore, when $N$ is sufficiently large, the ratio of the two elements in the output vector of a well-trained CMNet, as shown in (\ref{L-CMNet-EN}), is identical to the expression of the optimal LRT exploiting the perfect CSI, as defined in (\ref{T-EN}), i.e., the proposed CMNet method is equivalent to the optimal LRT detector.}

\section{Numerical Results}
In this section, we provide extensive simulation results to evaluate the performance of the proposed algorithm. Without special notes, we consider an AmBC system which consists of a single-antenna RF source, a single-antenna tag, and a 8-element ($M = 8$) multi-antenna reader{{\footnotemark}}\footnotetext{
{Considering the requirement of the coverage area and the adopted carrier frequency of the RFID reader \cite{van2018ambient, finkenzeller2010rfid}, we assume that $M=8$ antennas are equipped at the reader.}}. According to the  framework structure in Fig. 2, we set the length of the framework as $NT = 5000$, and the number of the pilots as $P = 10$, which is based on a practical protocol \cite{coleri2002channel}. To evaluate the BER performance, we compare the proposed CMNet method with the optimal LRT method \cite{kay1998fundamentals}, the classical ED method with perfect CSI\cite{qian2017semi}, and the SVM method \cite{hu2019machine}. {The hyperparameters of CMNet are given in Table \Rmnum{1}, where the optimizer used for training is Adam \cite{goodfellow2016deep}.
In addition, the numbers of examples in offline dataset and online dataset are $60,000$ and $2,000$, respectively.
The sizes of the input and the label of each example are shown in Table \Rmnum{1}.
The online dataset is generated by using the data augmentation technique \cite{shorten2019survey}, i.e., adding Gaussian noise variables with a distribution of $\mathcal{CN}(0,0.001)$ to the received $10$ pilot symbols to generate $2,000$ examples.}
{Note that although the proposed CMNet has large-size layers with large number of parameters, the online detection time can be greatly reduced through the parallelization of graphics processing unit (GPU) \cite{goodfellow2016deep}. For example, when we execute the proposed CMNet algorithm on a desktop computer with an i7-6700 3.4 GHz central processing unit (CPU) and a Nvidia GeForce GTX 1080 GPU, the online detection time of the CMNet algorithm is only 2.6 milliseconds, which is shorter than the coherence time of a general slow fading environment \cite{tse2005fundamentals} and is acceptable for the practical AmBC systems \cite{van2018ambient}.}
In addition, the SNRs of the simulation results are defined in (\ref{SNR}), and the relative coefficient is set as $\zeta = -20$ dB, as defined in (\ref{relative_SNR}).
{Each point in the simulation results is obtained by averaging over $10^6$ Monte Carlo realizations.}

{Fig. \ref{Fig_BER_SNR}(a) presents the BER curves versus SNRs for different algorithms when the ambient source transmits QPSK modulated signals. Note that the ED method presented in the simulations is not the same as the conventional ED method without CSI. Instead, it is a genie-aided ED method where the perfect CSI is available, denoted by ``ED with perfect CSI''.
In addition, the SVM method is an energy-based machine learning method, whose input is the signal energy received at each antenna of the reader.
The simulation results show that the BER performance of the SVM method is slightly worse than that of the ED method with perfect CSI.
The reason is that although the SVM method can exploit more distinguishable features to further improve the BER performance, it does not know the perfect CSI for detection and thus there is still a performance gap compared with the method of ``ED with perfect CSI''.}
In contrast to the conventional methods, the proposed CMNet method outperforms both the ED and SVM-based methods substantially, achieving almost the same BER performance as the optimal LRT method. For example, the proposed CMNet method achieves a SNR gain of 4 dB at BER$\,\approx10^{-2}$ compared with the traditional SVM-based method. The reason is that the SVM-based method only makes decisions based on the features capturing the characteristics of energy of received signals. In contrast, the proposed method makes decisions depending on the inherent features of the covariance matrix including rich distinguishable features without explicitly estimating the channels, which are acquired through a DTL approach.

\begin{figure}[t]
\centering
  \renewcommand{\arraystretch}{1.0}	
  \setlength{\tabcolsep}{1pt}
\begin{tabular}{c c}
\includegraphics[width=8cm,height=7cm]{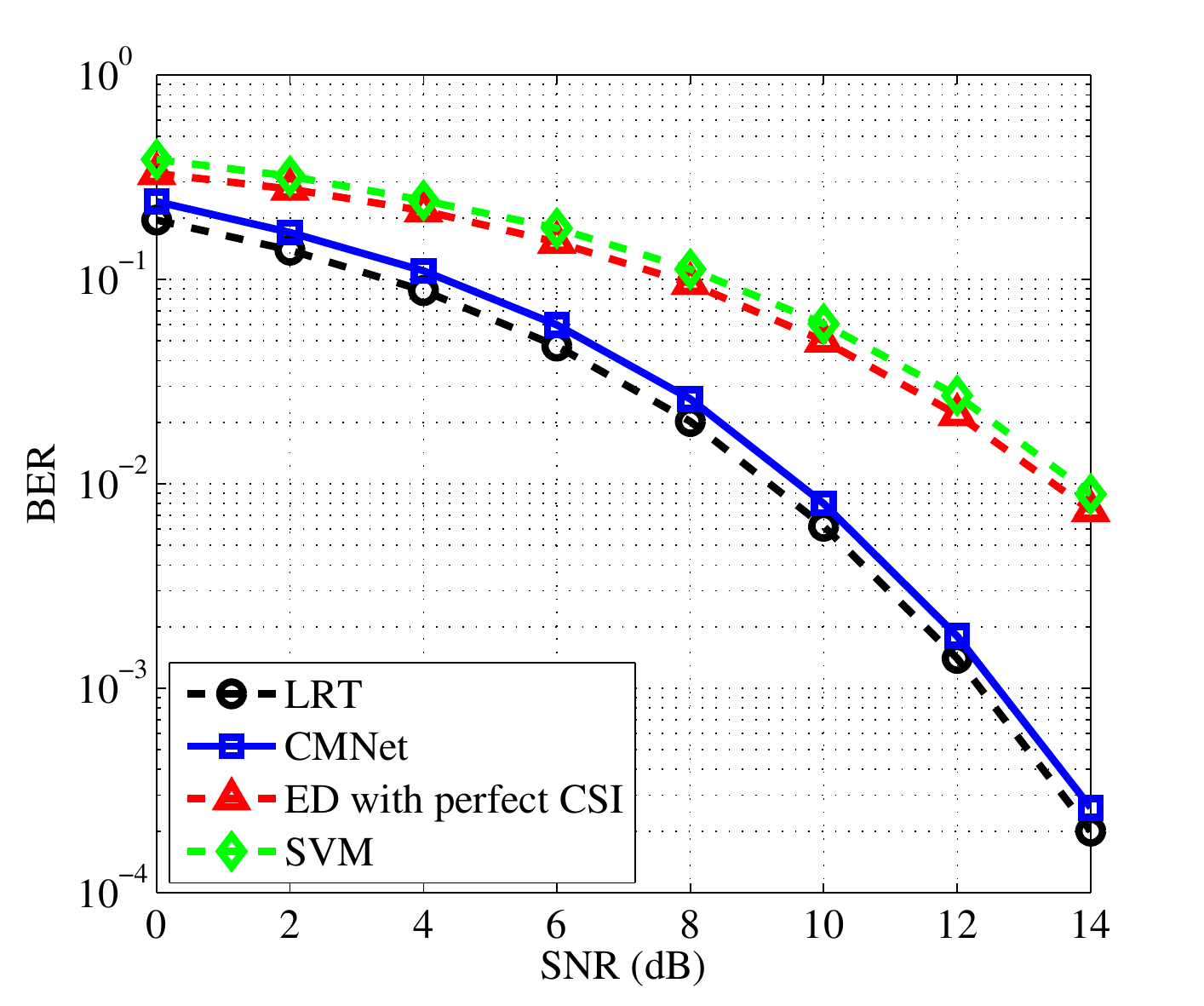}
&
\includegraphics[width=8cm,height=7cm]{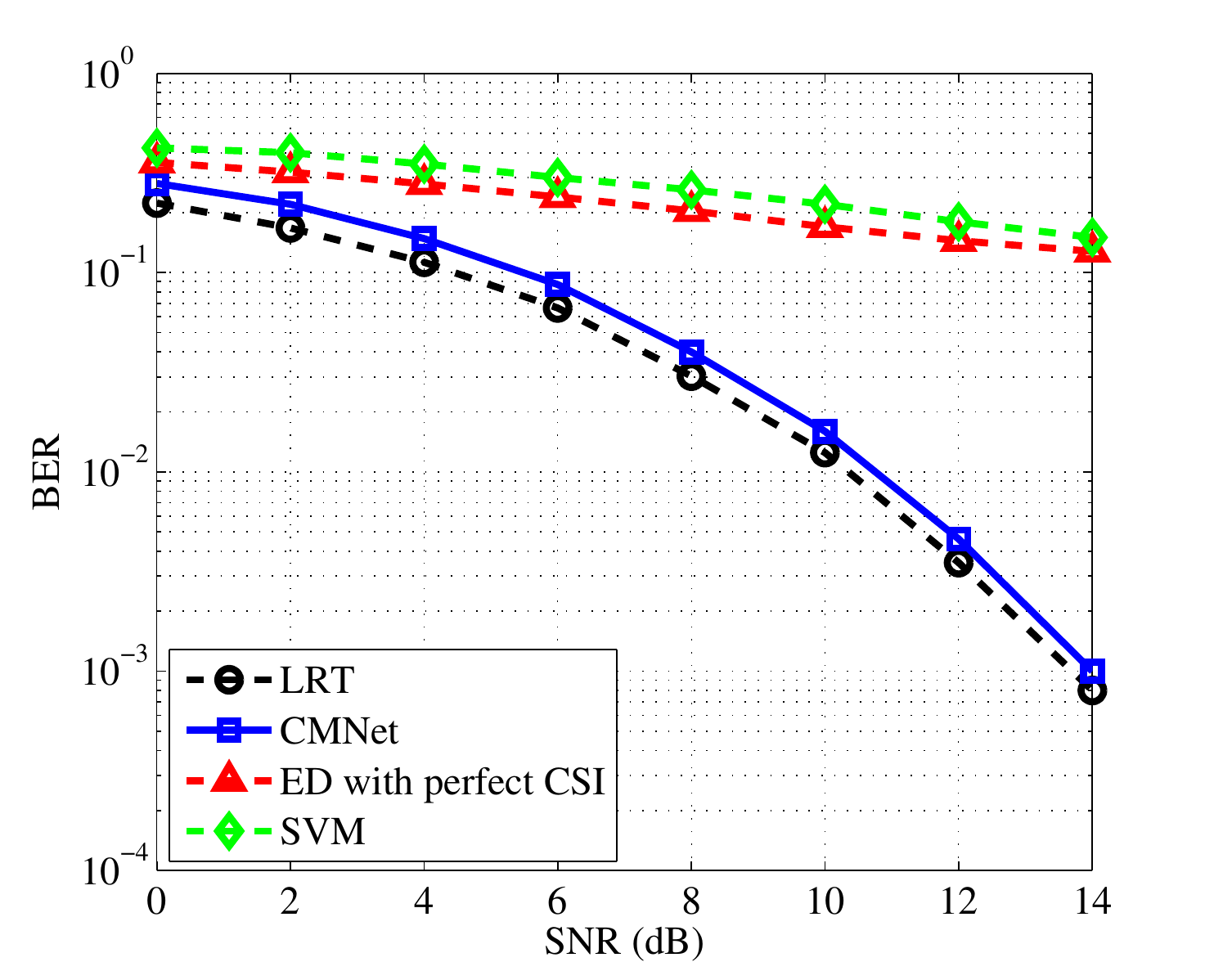} \vspace{-0.3cm}\\
{\scriptsize{(a) QPSK ambient source. }} &
{\scriptsize{(b) Complex Gaussian ambient source.}} \vspace{-0.2cm}\\
\end{tabular}
\caption{BER versus SNR with $M=8$, $N=20$, and $\zeta=-20~\mathrm{dB}$.}\label{Fig_BER_SNR} \vspace{-0.73cm}
\end{figure}

In addition, Fig. \ref{Fig_BER_SNR}(b) shows the BER-SNR curve under the complex Gaussian ambient source. {It is shown that the BER performance of all the presented algorithms under the complex Gaussian ambient source is slightly worse than that under the QPSK ambient source, as it is more difficult to distinguish the two hypotheses in the former case.} Similar to the results in Fig. \ref{Fig_BER_SNR}(a), our proposed CMNet also presents a satisfying BER performance which is almost the same as the performance of the optimal LRT method. Thus, the proposed method could achieve almost the optimal BER performance under both the QPSK and the complex Gaussian ambient sources.

\begin{figure}[t]
\centering
  \renewcommand{\arraystretch}{1.0}	
  \setlength{\tabcolsep}{1pt}
\begin{tabular}{c c}
\includegraphics[width=8cm,height=7cm]{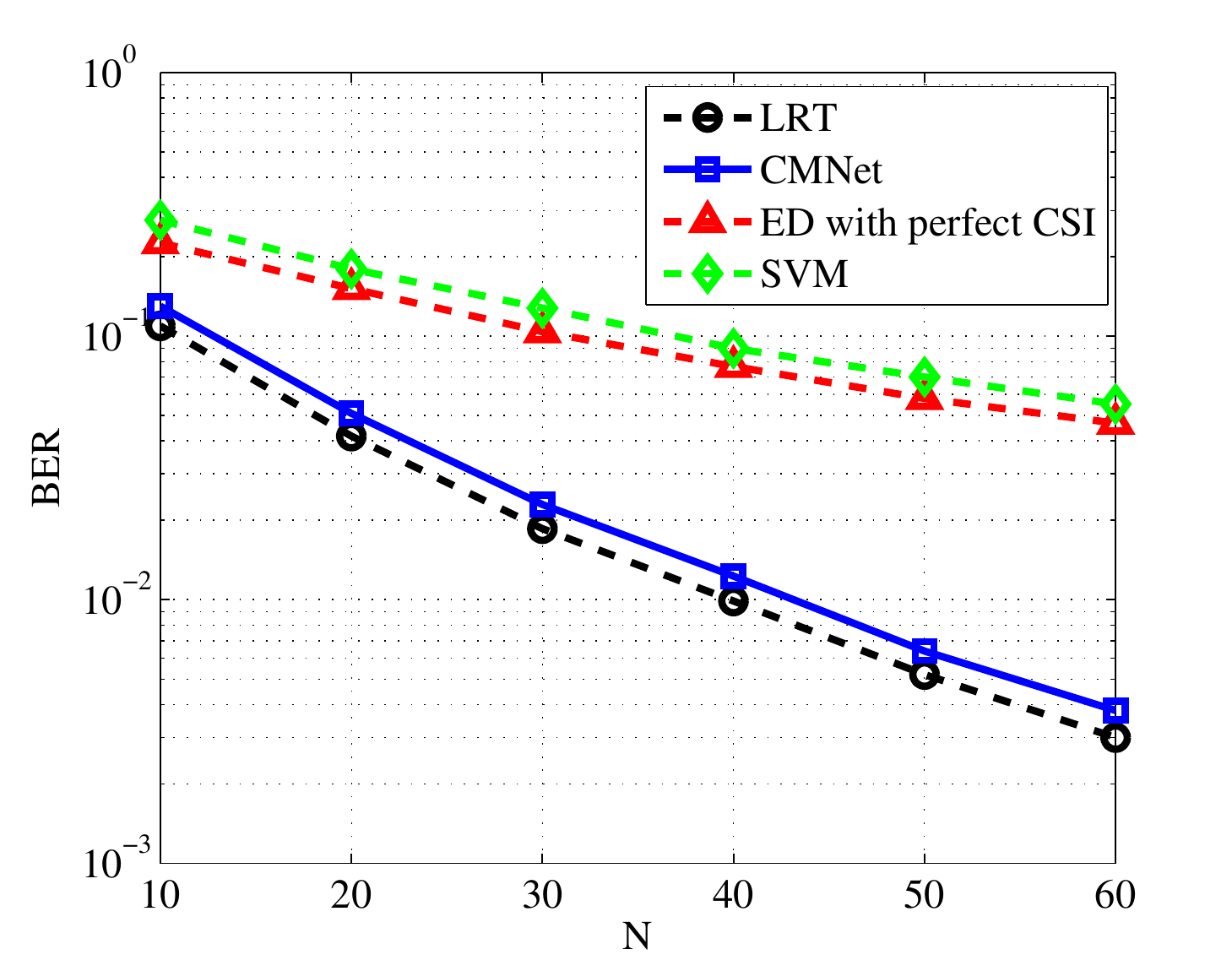}
&
\includegraphics[width=8cm,height=7cm]{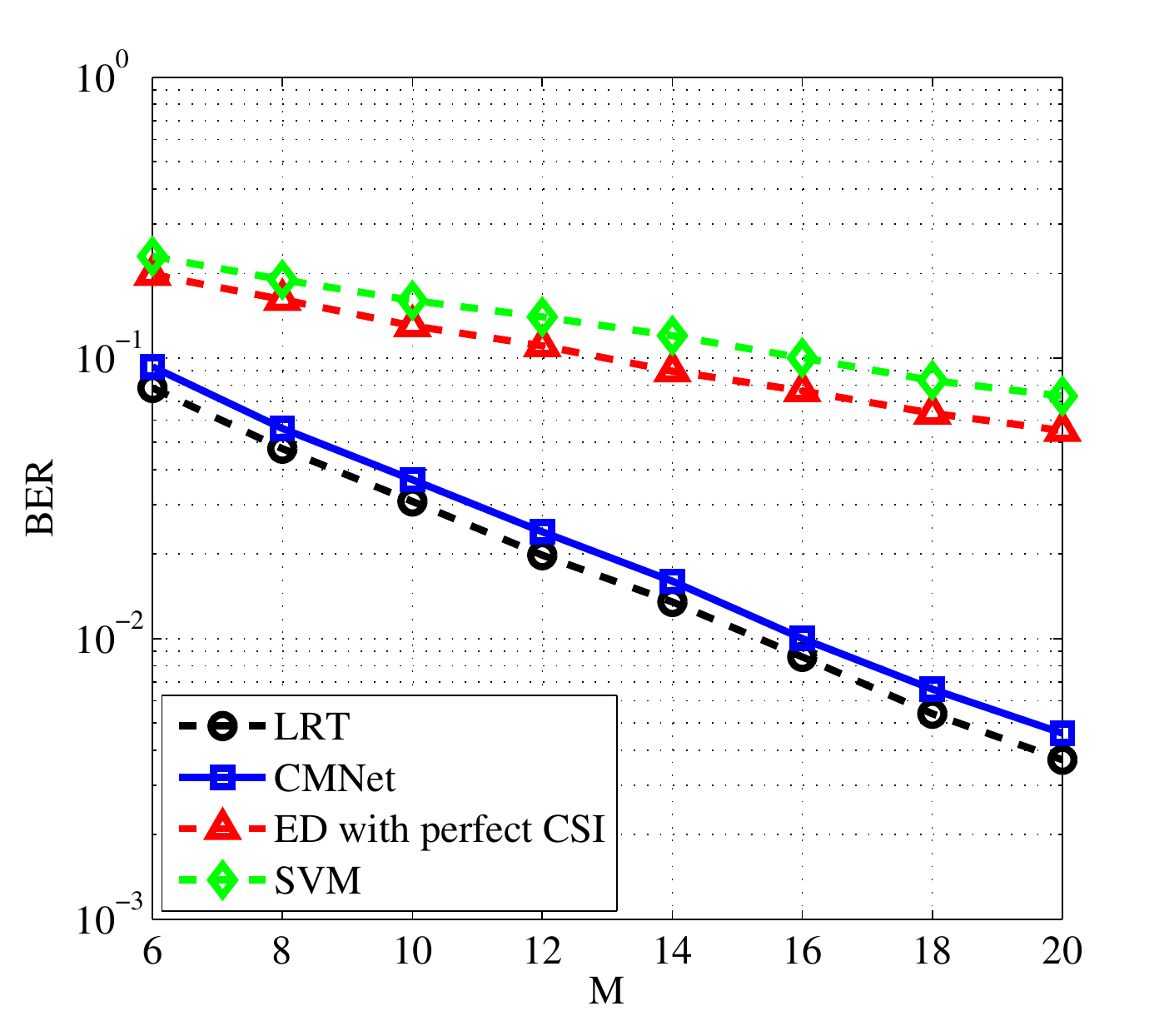} \vspace{-0.3cm}\\
{\scriptsize{(a) $M=8$, $\mathrm{SNR}=6~\mathrm{dB}$, and $\zeta = -20~\mathrm{dB} $. }} &
{\scriptsize{(b) $N=25$, $\mathrm{SNR}=5~\mathrm{dB}$, and $\zeta = -20~\mathrm{dB} $. }} \vspace{-0.2cm}\\
\end{tabular}
\caption{BER versus SNR under a QPSK ambient source.}\label{Fig_BER_MN}
\vspace{-0.73cm}
\end{figure}

Note that the sample covariance matrix heavily depends on the STR, $N$, and $M$. To evaluate the scalability of the proposed CMNet method, we fix the SNRs and vary the values of $N$ and $M$, presenting the curves of BER versus $N$ and BER versus $M$ in Fig. \ref{Fig_BER_MN}. From the results of Fig. \ref{Fig_BER_MN}(a), we can find that the BER of each detection algorithm decreases with an increasing $N$. Among all the detection algorithms, the BER performances of LRT and CMNet methods improve more dramatically than that of the ED and SVM-based algorithms.
Specifically, with the increasing of $N$, the BER of the proposed CMNet method scales with the same slope as the optimal LRT method, which demonstrates the scalability of the proposed method.  {Similar results can also be found in Fig. \ref{Fig_BER_MN}(b), which show that the proposed method is able to achieve outstanding BER performance under different numbers of antennas. This is because the proposed method can learn more distinguishable features from a larger scale input matrix to improve the accuracy of detection. Therefore, the proposed method is scalable for different STRs and numbers of antennas.}
{In addition, Fig. \ref{Fig_BER_MN}(b) shows that all algorithms perform closely when $M=6$ and the proposed CMNet achieves a significant performance gain when $M$ is large. This is because when $M$ is small, the differences between the two hypotheses become small, which hinders the signal detection in all the algorithms. In contrast, for a reasonably large $M$, the proposed CMNet can efficiently exploit the spatial degrees of freedom, which facilitates the signal detection.}

\begin{figure}[t]
\centering
  \renewcommand{\arraystretch}{1.0}	
  \setlength{\tabcolsep}{1pt}
\begin{tabular}{c c}
\includegraphics[width=8cm,height=7cm]{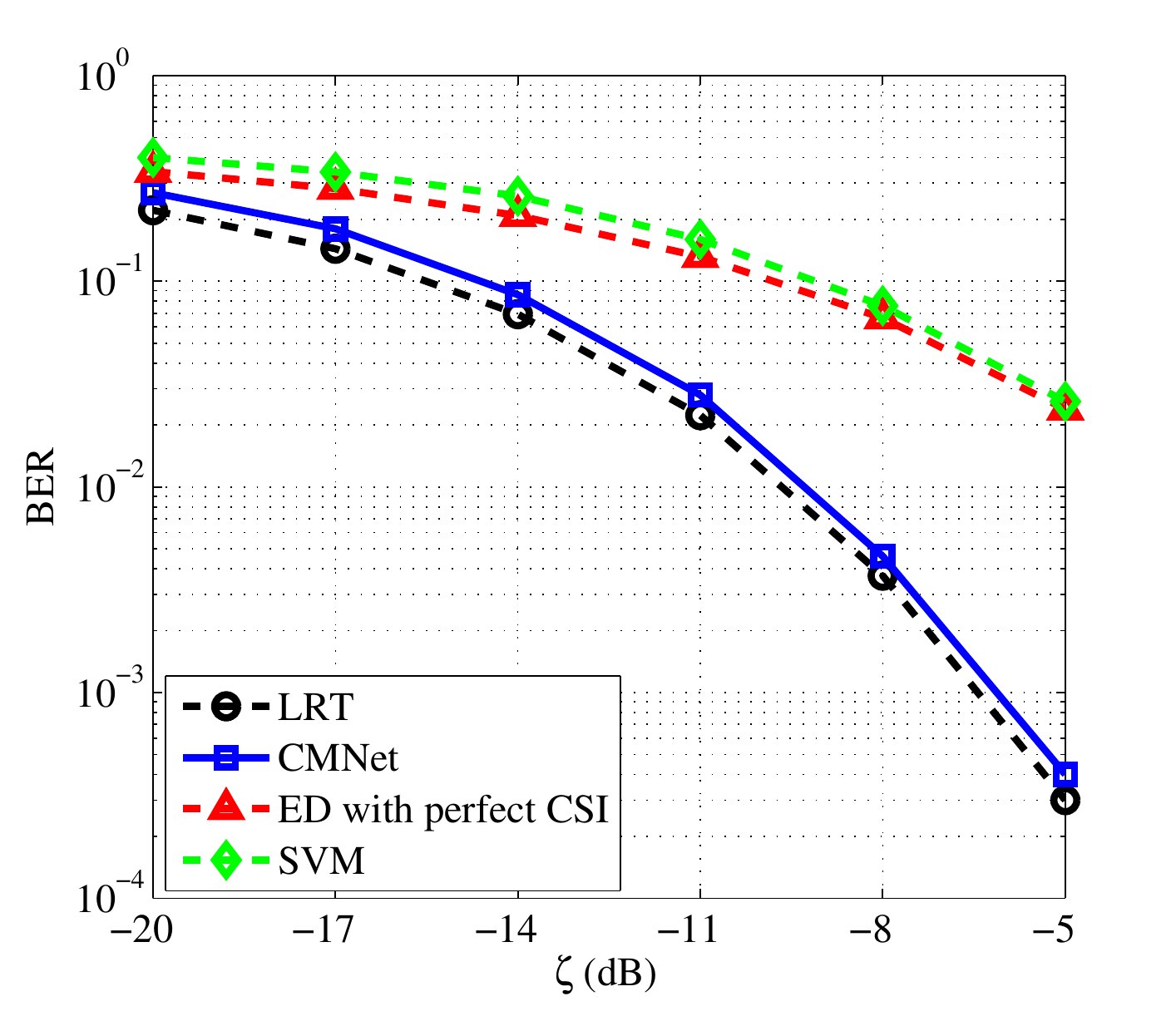}
&
\includegraphics[width=8cm,height=7cm]{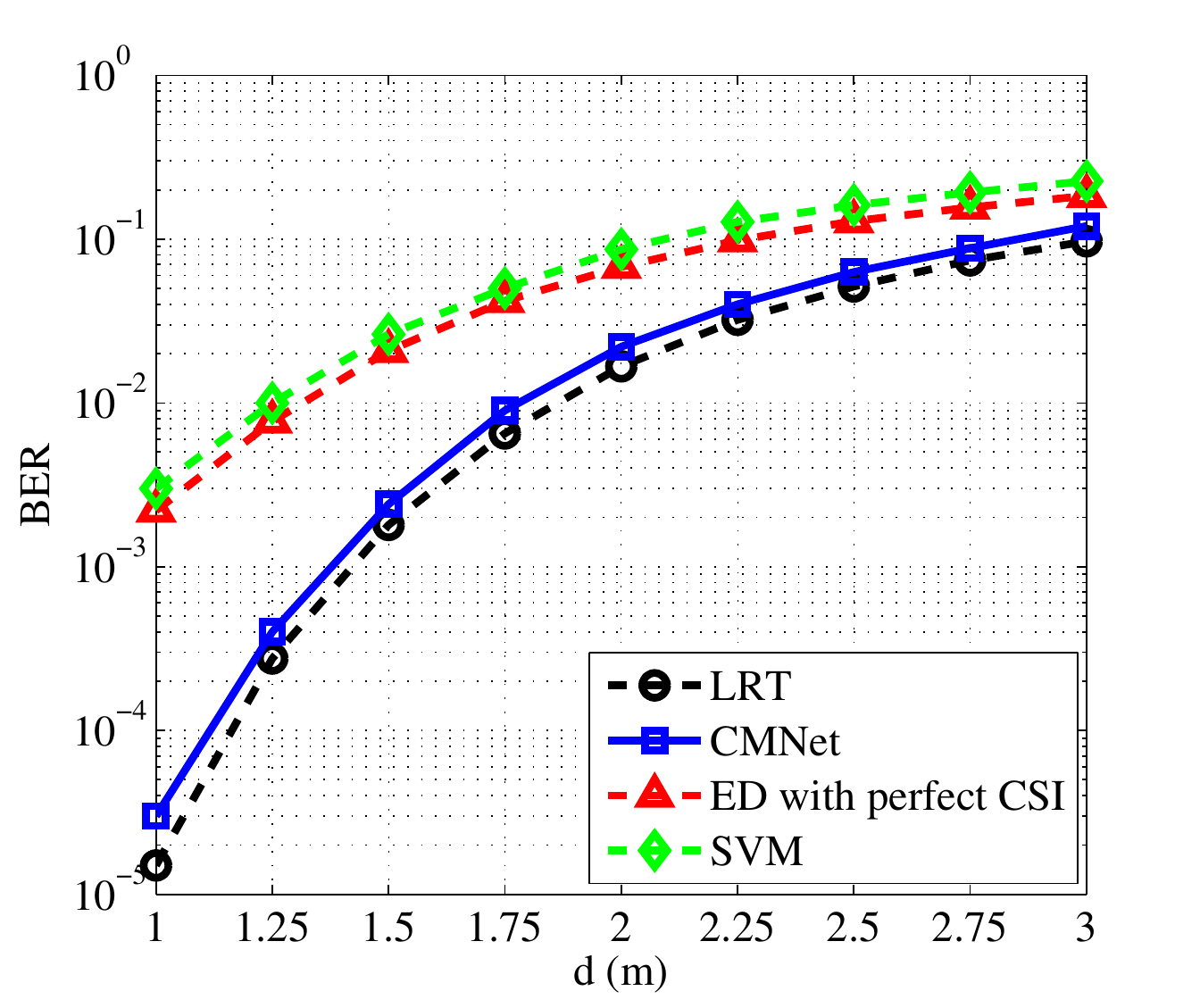} \vspace{-0.3cm}\\
{\scriptsize{(a) BER versus relative coefficient with $N=10$ and $\mathrm{SNR}=2~\mathrm{dB}$. }} &
{\scriptsize{(b) BER versus distance with $N=20$ and $\mathrm{SNR}_{\mathrm{tag}}=28~\mathrm{dB}$.}} \vspace{-0.2cm}\\
\end{tabular}
\caption{BER curves with different relative coefficients and tag-to-reader distances under a QPSK ambient source with $M=8$.}\label{Fig_BER_rlink}
\vspace{-0.73cm}
\end{figure}

{Finally, we turn to the study of the impact of reflection link on the system BER performance. As shown in Fig. \ref{Fig_BER_rlink}(a), a practical range of relative coefficient $\zeta \in [-20,-5]$ is investigated \cite{van2018ambient}. It is shown that the BER of each detection algorithm decreases with the increase of the relative coefficient $\zeta$, i.e., increasing the value of reflected coefficient contributes to the improvement of BER performance. This is because the improved strength of the reflected path makes the reader easier to distinguish the tag signals from the signals of the direct path.
In addition, although both the ED and SVM-based methods obtain some performance improvements by increasing the value of $\zeta$, the proposed CMNet method still outperforms both the SVM and the ED methods, achieving almost the same performance as the optimal LRT detector.}
{Note that the supported tag-to-reader distance is of great importance for practical implementation of AmBC systems. We then present the BER curves with different tag-to-reader distances in Fig. \ref{Fig_BER_rlink}(b).
In the simulations, a QPSK ambient RF source with a carrier frequency of $900~\mathrm{MHz}$ is adopted \cite{finkenzeller2010rfid} and the received SNR at the tag is set as $\mathrm{SNR}_{\mathrm{tag}} = 28~\mathrm{dB}$ \cite{van2018ambient}. In addition, a path loss model \cite{cho2010mimo} is introduced to characterize the large-scale fading of the tag-to-reader link, i.e., $\zeta = \beta(d/d_0)^{-\gamma}$, where $d$ is the tag-to-reader distance, $d_0=1~\mathrm{m}$ is the reference distance, $\gamma=2.7$ \cite{cho2010mimo} denotes the path loss exponent, and $\beta = (\lambda/(4\pi d_0))^2$ denotes the path loss of the signal with a wavelength of $\lambda$ at $d_0$.
It is shown from Fig. \ref{Fig_BER_rlink}(b) that a supported tag-to-reader distance of $2~\mathrm{m}$ is achieved by our proposed method with a BER requirement of $10^{-2}$, which is sufficient for the ambient backscatter communications in many practical scenarios \cite{van2018ambient}.
In addition, the BER performance of the proposed method approaches that of the optimal LRT method.
This is because the proposed method can adapt itself to different channel environments by the knowledge transfer from the source domain to the target domain.}

\section{Conclusions}
This paper studied the tag signal detection problem for AmBC systems adopting the DTL technology.
Firstly, we designed a universal DTL-based tag signal detection framework, which uses a DNN to implicitly extract the features of communication channels and directly recover the tag symbols.
Based on the established pre-trained DNN and a few pilots, a DTL-LRT was obtained through transfer learning, which enables the design of an effective detector. Furthermore, exploiting the advantages of the CNN's powerful capability in exploring features of data in a matrix form, we then designed a CMNet for the sample covariance matrix and proposed a CMNet-based detection algorithm. In particular, theoretical analysis of the proposed CNN-based method was provided correspondingly. Finally, simulation results showed that the proposed CMNet method can achieve a close-to-optimal performance without explicitly obtaining the CSI, despite the ambient source transmits modulated signals or complex Gaussian signals.


\bibliographystyle{ieeetr}

\setlength{\baselineskip}{10pt}

\bibliography{ReferenceSCI2}

\end{document}